\documentclass{tlp}

\usepackage{english}
\usepackage{amssymb}
\usepackage{latexsym}
\usepackage{epsfig}
\usepackage{calc}
\usepackage{curves}
\usepackage{varioref}

\newcommand\numlist{nl}
\newcommand\intlist{il}

\def\six{\mbox{SICStus}}
\def\block{{\tt block}}

\def\falls{\leftarrow}

\newcommand{\iatoms}{{\mathbf B}_P^{\it Inp}}
\def\onetom{\{1,\dots,m\}}
\def\oneton{\{1,\dots,n\}}

\newcommand{\step}{;}
\newcommand{\steps}{;\dots;}
\newcommand{\stopper}{\hspace*{0.5em} \hfill $\lhd$}

\def\ignore#1{}
\def\commentIgnore#1{}
\newcommand{\nat}{{\rm I \hspace{-0.2em} N}}
\newcommand{\perm}{permutation\ }
\newcommand{\vect}[1]{{\mathbf #1}} 
\newcommand{\dom}{{ dom}}
\newcommand{\ran}{{ ran}}
\newcommand{\vars}{{ vars}}

\newcommand{\inp}{{\it I}}
\newcommand{\out}{{\it O}}

\newcounter{xmiddle}
\newcounter{ymiddle}

\newcommand{\curl}[4]{%
\setcounter{xmiddle}{#1+#3}
\setcounter{xmiddle}{\thexmiddle/2}
\setcounter{ymiddle}{#2+#4}
\setcounter{ymiddle}{\theymiddle/2}
\qbezier[15](#1,#2)%
        (#1,\value{ymiddle})%
        (\value{xmiddle},\value{ymiddle})
\qbezier[15](\value{xmiddle},\value{ymiddle})%
        (#3,\value{ymiddle})%
        (#3,#4)}

\def\barray#1{\[\begin{array}{#1}}
\def\earray{\end{array}\]}

\def\newplaintheorem#1#2#3{%
\newtheorem{#1plain}[#3]{#2}
\newenvironment{#1}{\begin{#1plain}\rm}{\end{#1plain}}}

\newplaintheorem{example}{Example}{theo}
\newplaintheorem{defi}{Definition}{theo}
\newplaintheorem{regel}{Rule}{theo}
\newplaintheorem{coro}{Corollary}{theo}
\newplaintheorem{rmtheorem}{Theorem}{theo}
\newplaintheorem{rmlemma}{Lemma}{theo}
\newplaintheorem{propo}{Proposition}{theo}

\begin{document}
\bibliographystyle{tlp}

\title[Verifying Termination and Error-Freedom of Logic Programs] 
{Verifying Termination and Error-Freedom of Logic Programs 
with {\tt block} Declarations}

\author[J.--G. Smaus, P. M. Hill and A. M. King]
       {JAN--GEORG SMAUS\thanks{
Supported by \mbox{EPSRC} Grant No.~GR/K79635 and the ERCIM fellowship
       programme.}\\
CWI, \\
Kruislaan 413, 1098 SJ Amsterdam, The Netherlands\\
\email{jan.smaus@cwi.nl}\\
\and PATRICIA M.~HILL\\
School of Computer Studies, University of Leeds, \\
Leeds, LS2~9JT, United Kingdom,\\
\email{hill@scs.leeds.ac.uk} 
\and ANDY KING\\
University of Kent at Canterbury,\\
Canterbury, CT2~7NF, United Kingdom,\\
\email{a.m.king@ukc.ac.uk}
}

\maketitle

\begin{abstract}
We present verification methods for logic programs with delay
declarations. The verified properties are termination and freedom from
errors related to built-ins. Concerning termination, we present two
approaches. The first approach tries to eliminate the well-known
problem of {\em speculative} output bindings.  The second approach is
based on identifying the predicates for which the textual position of
an atom using this predicate is irrelevant with respect to
termination.  Three features are distinctive of this work: it allows
for predicates to be used in several modes; it shows that \block\
declarations, which are a very simple delay construct, are sufficient
to ensure the desired properties; it takes the selection rule into
account, assuming it to be as in most Prolog implementations.  The
methods can be used to verify existing programs and assist in writing
new programs.\\\\
{\bf Note:} This paper will be published in 
{\em Theory and Practice of Logic Programming}.
\end{abstract}

\section{Introduction}
The standard selection rule in logic programming states that the
leftmost atom in a query is selected in each derivation step. However,
there are some applications for which this rule is inappropriate,
e.g.~multiple modes, the test-and-generate paradigm~\cite{N92} or
parallel execution~\cite{AL95}.  To allow for more user-defined
control, several logic programming languages provide {\em delay
declarations}~\cite{goedel,sicstus}.  An atom in a query is selected
for resolution only if its arguments are instantiated to a specified
degree.  This is essential to ensure termination and to prevent
runtime errors produced by built-in predicates ({\em built-ins}).

In this paper we present methods for verifying programs with delay
declarations. We consider two aspects of verification: Programs should
terminate, and there should be no type or instantiation errors related
to the use of built-ins.

Three distinctive features of this work make its contribution:
(a) it is assumed that predicates may
run in more than one mode;
(b) we concentrate on \block\ declarations, which are a particularly
simple and efficient delay construct;
(c) the selection rule is taken into account. We now motivate these
features. 

(a) 
Allowing predicates to run in more than one mode is one
application of delay declarations. Although other
authors~\cite{AL95,N92} have not explicitly assumed multiple modes,
they mainly give examples where delay declarations are clearly used for
that purpose. Whether allowing multiple modes is a good approach or
whether it is better to generate multiple versions of each
predicate~\cite{mercury} is an ongoing discussion~\cite{newsletter}.
Our theory allows for multiple modes, but of course this does not
exclude other applications of delay declarations.

(b) 
The \block\ declarations declare that certain arguments of an atom
must be {\em non-variable} before that atom can be selected for
resolution. Insufficiently instantiated atoms are delayed. As
demonstrated in \six~\cite{sicstus}, \block\ declarations can be
efficiently implemented; the test whether arguments are non-variable
has a negligible impact on performance. Therefore such
constructs are the most frequently used delay declarations. Note that
most results in this paper also hold for other delay 
declarations considered in the literature. This is discussed in 
Sec.~\ref{discussion}.

(c) Termination may
critically depend on the selection rule, that is the rule which
determines, for a derivation, the order in which atoms are selected.
We assume that derivations are {\em left-based}. These are derivations
where (allowing for some exceptions concerning the execution order of
two literals woken up simultaneously) the leftmost selectable atom
is selected. This is intended to model derivations in the common
implementations of Prolog with \block\ declarations.
Other authors have avoided the issue by abstracting 
from a particular selection rule~\cite{AL95,L93}; considering
left-based selection rules on a heuristic basis~\cite{N92}; or making
the very restrictive assumption of {\em local selection
rules}~\cite{MT99}.

The main contribution concerns termination. We have isolated some of
the causes of non-termination that are related to the use of delay
declarations and identified conditions for programs to avoid those
causes. These conditions can easily be checked at compile-time.  The
termination problem for a program with delay declarations is then
translated to the same problem for a corresponding program executed
left-to-right.  It is assumed that, for the corresponding program,
termination can be shown using some existing
technique~\cite{A97,DD94,EBC99}.

One previously studied cause of non-termination 
associated with delay declarations is
{\em speculative output bindings}~\cite{N92}. 
These are bindings made before it is known that a solution exists. 
We present two complementing methods for dealing with this problem 
and thus proving (or ensuring) termination.
Which method must be applied will depend on the program
and on the mode being considered.  
The first
method exploits that a program does not {\em use} any speculative
bindings, by ensuring that no atom ever delays. 
The second method exploits that a
program does not {\em make} any speculative bindings.  

However, these two methods are quite limited. 
As an alternative approach to the termination problem,
we identify certain predicates that may loop when
called with {\em insufficient} 
(that is, non-variable but still insufficiently instantiated) input. 
 For instance, with the predicate $\tt permute/2$ where the second
argument is input, the query 
\verb@permute(A,[1|B])@ has insufficient input and loops.\footnote{
The program for $\tt permute/2$ is given in Fig.~\ref{permute-ok-fig}.}
However, the query 
\verb@permute(A,[1,2])@ has sufficient input and terminates. 
The idea for proving termination is that, for such predicates, 
calls with insufficient input must never arise. 
This can be ensured by appropriate ordering of atoms in the clause
bodies. This actually works in several modes provided not too many
predicates have this undesirable property.

Our work on built-ins focuses on {\em arithmetic} built-ins. By
exploiting the fact that for numbers, being non-variable implies being
ground, we show how both {\em instantiation} and {\em type} errors can
be prevented.

Finally, we consider two other issues related to delay declarations.
First, we identify conditions so that certain \block\ 
declarations can be omitted without affecting the runtime behaviour.
Secondly, to verify programs with delay declarations, it is often
necessary to impose a restriction on the modes that 
forbids tests for identity between the input arguments of an atom. We
explain how this rather severe restriction is related to the use of
delay declarations and how it can be weakened.

This paper is organised as follows. The next section
defines some essential concepts and notations. 
Section~\ref{easy} introduces four concepts of ``modedness'' and 
``typedness'' that are needed later.
Section~\ref{termination-a}, 
which is based on previously published work~\cite{HKS-error}, 
presents the first approach to the termination problem.
Section~\ref{termination-b}, 
which is also based on previously published work~\cite{HKS-term}, 
presents the second approach.
Section~\ref{error-sec} is about errors related to built-ins.
Section~\ref{improve-sec} considers ways of simplifying 
the \block\ declarations.
Section~\ref{related} investigates related work.
 Section~\ref{discussion} concludes with a summary and a 
look at ongoing and future work. 

\section{Essential Concepts and Notations} \label{prelims}

\subsection{Standard Notions}\label{standard-subsec}
We base the notation on~\cite{AL95,L87}.
For the examples we use \six\ notation~\cite{sicstus}.
A term $u$ occurs {\bf directly} in a vector of terms $\vect{t}$ if $u$ is one of the
terms of~$\vect{t}$. 
(For example, $a$ occurs directly in $(a,b)$ but not in $(f(a),b)$.)
We also say that $u$ {\bf fills a position in} $\vect{t}$.
To refer to the predicate symbol of an atom, 
we say that an atom $p(\dots)$ is an atom {\bf using} $p$.
The set of variables in a syntactic object $o$ is denoted by
$\vars(o)$. 
A syntactic object is {\bf linear} if every variable occurs in it at
most once.
Otherwise it is {\bf non-linear}.
A {\bf flat} term is a variable or a term $f(x_1,\dots,x_n)$, 
where
$n \geq 0$ and the $x_i$ are distinct variables.
The {\bf domain} of a substitution $\sigma$  is
$\dom(\sigma) = \{ x \mid x \sigma \neq x \}$. 
The variables in the range of $\sigma$ are denoted as
$\ran(\sigma)=  \{ y \mid y \in \vars(x \sigma), y \neq x \}$.
\ignore{
The restriction of a substitution $\theta$ onto the variables in a
syntactic object $o$ is denoted as $\theta | o$.}

A {\bf query} is a finite sequence of atoms.
Atoms are denoted by $a$, $b$, $h$, queries by $B$, $F$, $H$, $Q$, $R$. 
Sometimes we say ``atom'' instead of ``query consisting of an atom''.
A {\bf derivation step} for a program $P$ is a pair 
$\langle Q,\theta        \rangle;
 \langle R,\theta \sigma \rangle$,  
where $Q = Q_1,a,Q_2$ and $R = Q_1,B,Q_2$ are queries; 
$\theta$ is a substitution; 
$a$ an atom;
$h \falls B$ a variant of a clause in $P$,
renamed apart from $Q\theta$,  
and $\sigma$ the  most general unifier (MGU) of $a \theta$ and $h$.
We call $a \theta$ (or $a$)\footnote{
Whether or not the substitution has been applied is always clear from the context.
} 
the {\bf selected atom}
and $R \theta \sigma$ the 
{\bf resolvent} of $Q \theta$ and $h \falls B$.

A {\bf derivation $\xi$ for} a program $P$ is a sequence 
$\langle Q_0, \theta_0 \rangle; 
 \langle Q_1, \theta_1 \rangle; \dots$ where 
each pair 
$\langle Q_i, \theta_i \rangle; 
 \langle Q_{i+1}, \theta_{i+1} \rangle$ 
in $\xi$ is a derivation step for $P$.
Alternatively, we also say that $\xi$ is a 
{\bf derivation of} $P \cup \{Q_0 \theta_0\}$.
We also denote $\xi$ by
$Q_0 \theta_0; 
 Q_1 \theta_1; \dots$.
A derivation is an {\bf LD-derivation} if the selected atom is
always the leftmost atom in a query.

\addtocounter{footnote}{-1}
If 
$
F,a,H;\, (F, B, H) \theta
$
is a step in 
a derivation, then each atom in $B \theta$ 
(or $B$)\footnotemark \addtocounter{footnote}{-1}
is a {\bf direct descendant}
of $a$, and $b\theta$ (or $b$)\footnotemark \
is a {\bf direct descendant} of $b$ 
for all $b$ in $F,H$.
We say that $b$ is a {\bf descendant of} $a$, or 
$a$ is an {\bf ancestor of} $b$, if $(b,a)$ is in the
reflexive, transitive closure of the relation {\em is a direct
  descendant}.
The descendants of a {\em set} of atoms are defined in the obvious way.
Consider a derivation
$Q_0 \steps Q_i\steps Q_j \step Q_{j+1}; \dots$.
We call $Q_j \step Q_{j+1}$ an  {\bf $a$-step} if $a$ is an atom
in $Q_i$ ($i\leq j$) and the selected atom in $Q_j \step Q_{j+1}$ 
is a descendant of $a$. 

\subsection{Modes}\label{modes-subsec}
For a predicate $p/n$, a {\bf mode} is an atom $p(m_1,\dots,m_n)$,
where $m_i \in \{{\tt \inp,\out} \}$ for $i \in \oneton$.  Positions
with $\inp$ are called {\bf input positions}, and positions with
$\out$ are called {\bf output positions} of $p$.  A {\bf mode of a
  program} is a set of modes, one mode for each of its predicates.  
An atom written as $p(\vect{s},\vect{t})$
means: $\vect{s}$ and $\vect{t}$ are the vectors of terms 
filling the input and output positions of $p$, respectively.

An atom $p(\vect{s},\vect{t})$ is
{\bf input-linear} if $\vect{s}$ is linear.
A clause is {\bf input-linear} if its head is input-linear.  
A program is {\bf input-linear}
if all of its clauses are
input-linear and it contains no uses of  
$\mbox{\tt =}(\inp,\inp)$.\footnote{Conceptually, one can think of 
each program containing the fact clause $\tt X = X.$}

We claim that the techniques we describe are suitable for programs 
that can run
in several modes. Throughout most of the presentation, this is not
explicit, since we always consider one mode at a time. Therefore,
whenever we refer to the input and output positions, this is always
with respect to one particular mode. However, we will see in several
examples that {\em one single} program can be ``mode correct'', in a
well-defined sense, with respect to several different modes. 
In particular, {\em one single} delay declaration for a predicate can
allow for this predicate to be used in different modes. 

This is different from the assumption made by some
authors~\cite{AE93,AL95,EBC99,N92} that if a predicate is to be used
in several modes, then multiple (renamed) versions of this predicate
should be introduced, which may differ concerning the delay
declarations and the order of atoms in clause bodies.

Note that our notion of modes could easily be generalised further by
assigning a mode to predicate {\em occurrences} rather than
predicates~\cite{smaus-thesis}. 

\subsection{Types}\label{types-subsec}
A {\bf type} is a set of terms closed under instantiation~\cite{AL95,B96}.
{\em The} {\bf variable type} is the type that contains variables
and hence, as it is closed under instantiation, all terms.
Any other type is a {\bf non-variable type}.
A type is 
a {\bf ground type} if it contains only ground terms.
A type is 
a {\bf constant type} if it is a ground type that 
contains only (possibly infinitely many) constants.
In the examples, we use the following types:
$any$ is the variable type, 
$list$ the non-variable type of (nil-terminated) lists,
$int$ the constant type of integers, 
$\intlist$ the ground type of integer lists, 
$num$ the constant type of numbers, 
$\numlist$ the ground type of number lists, 
and finally,
$tree$ is the non-variable type defined by the context-free grammar
$\{ tree \rightarrow {\tt leaf}; 
tree \rightarrow {\tt node}(tree,any,tree) \}$. 

We write $t:T$ for ``$t$ is in type $T$''.
We use $\vect{S}$, $\vect{T}$ to denote vectors of types,
and write 
$\models \vect{s}: \vect{S} \Rightarrow \vect{t}: \vect{T}$
if for all substitutions $\sigma$, 
$\vect{s}\sigma : \vect{S}$ implies $\vect{t}\sigma: \vect{T}$.
It is assumed that each argument position of each predicate $p/n$
has a type associated with it. 
These types are indicated by writing the atom $p(T_1,\ldots,T_n)$ 
where $T_1,\dots,T_n$ are types.
The {\bf type of a program} $P$ is a set of such atoms, one for each
predicate defined in $P$.
An atom is {\bf correctly typed} in a position if the term
filling this position has the type that is associated with this position. 
A term $t$ is  {\bf type-consistent with respect to}~$T$~\cite{DM98} if
there is a substitution $\theta$ such that $t \theta: T$.
A term $t$ occurring in an atom in some position
is {\bf type-consistent} if it is type-consistent with respect to the type of that
position. 

\subsection{\block\ Declarations}
A \block\ declaration~\cite{sicstus} for a predicate $p/n$ is a
(possibly empty) set of atoms each of which has the form
$p(b_1,\dots,b_n)$, where $b_i \in \{\mbox{\tt ?},\mbox{\tt -} \}$ for $i \in
\oneton$. 
A {\bf program} consists of a set of clauses and a 
set of \block\ declarations, 
one for each predicate defined by the clauses.
If $P$ is a program, an atom $p(t_1,\dots,t_n)$ is 
{\bf blocked in $P$} if there is an atom $p(b_1,\dots,b_n)$ in the \block\
declaration for $p$ such that for all 
 $i \in \oneton$ with $b_i = \mbox{\tt -}$, 
we have that $t_i$ is variable. 
An atom is 
{\bf selectable in $P$} if it is not blocked in $P$.

\begin{example}\label{block-ex}
Consider a program containing the \block\ declaration
\begin{verbatim}
:- block append(-,?,-), append(?,-,-).
\end{verbatim}
Then atoms
$\tt append(X,Y,Z)$, 
$\tt append([1|X],Y,Z)$, and
$\tt append(X,[2|Y],Z)$ are blocked in $P$, whereas atoms
$\tt append([1|X],[2|Y],Z)$,
$\tt append(X,Y,[1|Z])$ and
$\tt append(X,[2|Y],[1|Z])$ are selectable in $P$.
\end{example}

\noindent
Note that equivalent delay constructs are provided in several logic
programming languages, although there may be differences in the
syntax.

A {\bf delay-respecting derivation} for a program $P$
 is a derivation where the
selected atom is always selectable in $P$. 
We say that it {\bf flounders} if it ends with
a non-empty query where no atom is selectable. 

\subsection{Left-based Derivations}
\label{block-derivation-subsec}
We now formalise the sort of derivations that arise in practice using
almost any existing Prolog implementations. 
Some authors have considered a selection rule
  stating that in each derivation step, the leftmost selectable atom
  is selected~\cite{AL95,B96,N92}. We are not aware of an existing
  language that uses this selection rule, contradicting Boye's
  claim~\cite[(1996, page 123)]{B96} that several modern Prolog 
implementations and even 
G\"odel~\cite{goedel} use this selection rule.
 In fact, Prolog implementations do not
  usually guarantee the order in which two simultaneously woken atoms
  are selected.

\begin{defi}[left-based derivation]
\label{left-based-def}
Consider a delay-respecting derivation 
$Q_0 \steps Q_i; \dots$, where 
$Q_i = R_1,R_2$, and $R_1$ contains no selectable atom. 
Then every descendant of every atom in $R_1$ is {\bf waiting}.
A delay-respecting derivation $Q_0 \step Q_1 \dots$ is {\bf left-based} if for
each step $Q_i \step Q_{i+1}$, the selected atom is either
waiting in $Q_i$, or it is the leftmost selectable atom in $Q_i$.
\end{defi}

\begin{example} \label{left-based-ex}
Consider the following program:
{\small
\begin{verbatim}
:- block a(-).           :- block b(-)
a(1).                    b(X) :- b2(X).

c(1).                    b2(1).                   d.
\end{verbatim}
}
The following is a left-based derivation. 
Waiting atoms are underlined.
\[
\tt \underline{a(X)}, 
    \underline{b(X)}, c(X), d \step \quad
\tt \underline{a(1)}, 
    \underline{b(1)}, d \step \quad
\tt \underline{a(1)}, 
    \underline{b2(1)}, d \step \quad
\tt \underline{a(1)}, 
    d \step \quad
\tt d \step \quad
\Box.
\]
Note that $\tt b(1)$ and $\tt b2(1)$ are waiting and selectable,
and therefore
they can be selected although there is the selectable atom $\tt a(1)$ to
the left.
\stopper \end{example}

\noindent
We do not believe that it would be useful or practical to try to
specify the selection rule precisely, but from our research, it appears
that derivations in most Prolog implementations are
left-based. 

Note that the definition of left-based derivations for a program and query
depends both on the textual order of the atoms in the query and
clauses and on the \block\ declarations.
In order to maintain the textual order while considering 
different orders of selection of atoms, 
it is often useful to associate, with a query,
a permutation $\pi$ of the atoms.

Let $\pi$ be a permutation on $\oneton$. 
We assume that $\pi(i) = i$ for $i \notin
\oneton$.  In examples, $\pi$ is written as 
$\langle \pi(1), \dots,\pi(n) \rangle$.  We write
$\pi(o_1,\dots,o_n)$ for the application of $\pi$ to
the sequence $o_1,\dots,o_n$, that is
$o_{\pi^{-1}(1)},\dots,o_{\pi^{-1}(n)}$.

\section{Correctness Conditions for Verification} \label{easy}

Apt and Luitjes~\shortcite{AL95} consider three correctness conditions 
for programs: {\em nicely moded}, {\em well typed}, and {\em simply
moded}. 
Apt~\shortcite{A97} and Boye~\shortcite{B96} propose a generalisation 
of these
conditions that allows for permutations of the atoms in
each query.
Such correctness conditions have been used for various
verification purposes: occur-check freedom, flounder freedom, freedom
from errors related to built-ins~\cite{AL95}, freedom from
failure~\cite{BC98}, and termination~\cite{EBC99}.  In this section we
introduce four such correctness conditions and show some important
statements about them.  The correctness conditions will then be used
throughout the paper.

The idea of these correctness conditions is that in a query, every
piece of data is produced (output) before it is consumed (input), and
every piece of data is produced only once. The definitions of these
conditions have usually been aimed at LD-derivations, which means that
an output occurrence of a variable must always be to the left of any
input occurrence of that variable.

\subsection{Permutation Nicely Moded Programs}\label{nicely-subsec}
In a {\em nicely moded} query,
a variable occurring in an input position does not occur 
later in an output position, and each variable in an output
position occurs only once. We generalise this to 
{\em permutation nicely moded}.
Note that the use of the letters $s$ and $t$ is reversed for clause
heads. We believe that this notation naturally reflects the data flow
within a clause. This will become apparent in Def.~\ref{well-typed}. 

\begin{defi}[\perm nicely moded] \label{nicely-moded}
Let $Q = p_1(\vect{s}_1,\vect{t}_1),\dots,
p_n(\vect{s}_n,\vect{t}_n)$ be a query
  and
$\pi$ a permutation on $\oneton$.
Then $Q$ is 
{\bf $\pi$-nicely moded} if $\vect{t}_1,\dots,\vect{t}_n$ is a linear
 vector of terms and for all $i \in \oneton$
\[
\vars(\vect{s}_i) \cap \bigcup_{\pi(i) \leq \pi(j) \leq n} \vars(\vect{t}_j)
= \emptyset. 
\]

The query $\pi(Q)$
is a
{\bf nicely moded query corresponding to $Q$}.

The clause 
$C = p(\vect{t}_0,\vect{s}_{n+1}) \leftarrow Q$
is {\bf $\pi$-nicely moded} if 
$Q$ 
is $\pi$-nicely moded and 
\[
\vars(\vect{t}_0) \cap 
\bigcup_{j=1}^{n} \vars(\vect{t}_j) = \emptyset.
\]
The clause 
$p(\vect{t}_0,\vect{s}_{n+1}) \leftarrow \pi(Q)$
is a {\bf nicely moded clause corresponding to $C$}.

A query (clause) is {\bf \perm nicely moded} if it is $\pi$-nicely
moded for some $\pi$. 
A program $P$ is {\bf \perm nicely moded} if all of its clauses are. 
A {\bf nicely moded program corresponding to $P$} is a program
obtained from $P$ by replacing every clause $C$ in $P$ with a nicely
moded clause corresponding to $C$.
\stopper \end{defi}

\noindent
In Lemma~\ref{nicely-moded-persistence}, on which many results of this
paper depend, we require a program not only to be permutation nicely
moded, but also input-linear (see Subsec.~\ref{modes-subsec}).

\begin{example} \label{permute-loops} 
The program in Fig.~\ref{permute-loops-fig}
is nicely moded and input-linear in mode $M_1$.\footnote{
For convenient reference,
the modes are included in the figure. Also, the program contains 
{\tt block} declarations. We will refer to those later; they should be
ignored for the moment.} 
In mode $M_2$
it is \perm nicely moded and input-linear.
In particular, the second clause for {\tt permute} is 
$\langle 2,1 \rangle$-nicely moded.
In ``test mode'', that is, 
$\{ {\tt permute(\inp,\inp)},$ 
${\tt delete(\inp,\inp,\out)}\}$,  it
is \perm nicely moded, but not input-linear, because the first clause for 
{\tt delete} is not input-linear. 
\stopper \end{example}

\begin{figure}
\figrule
\begin{center}
\begin{minipage}[t]{6cm}
\small
\begin{verbatim}
:- block permute(-,-).
permute([],[]).
permute([U|X],Y) :- 
  permute(X,Z),
  delete(U,Y,Z).
\end{verbatim}
\end{minipage}
\begin{minipage}[t]{6cm}
\small
\begin{verbatim}
:- block delete(?,-,-).
delete(X,[X|Z],Z).
delete(X,[U|Y],[U|Z]) :- 
  delete(X,Y,Z).
\end{verbatim}
\end{minipage}
\end{center}

\vspace*{2ex}
$M_1 = \{ {\tt permute(\inp,\out)},
{\tt delete(\inp,\out,\inp)}\}$\\
$M_2 = \{ {\tt permute(\out,\inp)}, 
{\tt delete(\out,\inp,\out)}\}$
\caption{The {\tt permute} program \label{permute-loops-fig}}
\figrule
\end{figure}

\noindent
We show that there is a persistence property for \perm nicely-moded\-ness 
similar to that for nicely-modedness~\cite{AL95}.

\begin{rmlemma} \label{nicely-moded-persistence}
Let $Q =  a_1,\dots,a_n$ be a $\pi$-nicely moded query and 
$C = h \leftarrow b_1,\dots,b_m$ be a $\rho$-nicely moded, input-linear clause 
where $\vars(Q) \cap \vars(C) = \emptyset$.
Suppose for some $k \in \oneton$, $h$ and $a_{k}$ are unifiable. 
Then the resolvent of $Q$ and $C$ with selected atom $a_k$ is  
$\varrho$-nicely moded, where  
the {\bf derived permutation} $\varrho$ on $\{1,\dots, n+m-1\}$ is
defined by $\varrho(i)=$

\[
\!
\left\{
 \begin{array}{ll}
\pi(i) & \quad
\mbox{if} \ i < k,                                    \; \pi(i) < \pi(k)\\
\pi(i) + m - 1 & \quad
\mbox{if} \ i < k,                                    \; \pi(i) > \pi(k)\\
\pi(k) + \rho(i-k+1) -1 & \quad
\mbox{if} \ k \leq i < k+m \\
\pi(i-m+1) & \quad
\mbox{if} \ k+m \leq i < n+m,                    \; \pi(i-m+1) < \pi(k) \\
\pi(i-m+1)+m-1 & \quad
\mbox{if} \ k+m \leq i < n+m,                    \; \pi(i-m+1) > \pi(k).
\end{array}
\right.
\]
\end{rmlemma}

\begin{proof} 
Let $\theta$ be the MGU of $h$ and $a_{k}$.
By Def.~\ref{nicely-moded}, we have that
$a_{\pi^{-1}(1)},\dots,a_{\pi^{-1}(n)}$ and
$h \leftarrow b_{\rho^{-1}(1)},\dots,b_{\rho^{-1}(m)}$ are 
nicely moded and $h$ is input-linear. Thus by Lemma~11~\cite{AL95}
\[
(a_{\pi^{-1}(1)},\dots,a_{\pi^{-1}(\pi(k)-1)},
b_{\rho^{-1}(1)},\dots,b_{\rho^{-1}(m)},
a_{\pi^{-1}(\pi(k)+1)},\dots,a_{\pi^{-1}(n)}) 
\, \theta
\]
is nicely moded, and hence
$
(a_1,\dots,a_{k-1},b_{1},\dots,b_{m},a_{k+1},\dots,a_{n}) \,
\theta$
is 
\mbox{$\varrho$-nicely} moded.
 \end{proof}

\noindent
Figure~\ref{order-illus} illustrates
 $\varrho$ when $Q = a_1,a_2,a_3,a_4\,$, 
$\pi = \langle 4,3,1,2 \rangle\,$, 
$C = h \falls b_1,b_2\,$, 
$\rho = \langle 2,1 \rangle\,$, and $k = 2$. 
Thus   $\varrho = \langle 5,4,3,1,2 \rangle$.
Observe that, at each step of a derivation, 
the relative order of atoms given by the derived permutation is preserved. 
By a
straightforward induction on the length of a derivation, using the definition of $\varrho$ for the base case, we obtain the following corollary.

\begin{coro} \label{mode-order}
Let $P$ be a permutation nicely moded, input-linear program,
$Q = a_1,\dots,a_n$ be a $\pi$-nicely
moded query and 
$i,j \in \oneton$ such that $\pi(i) < \pi(j)$. Let $Q;\dots;R$ be a
derivation for $P$ and suppose $R = b_1,\dots,b_m$ is $\rho$-nicely
moded. If for some $k,l \in \onetom$, $b_k$ is a descendant of $a_i$
and $b_l$ is a descendant of $a_j$, then $\rho(k) < \rho(l)$.
\end{coro}

\setlength{\unitlength}{0.42cm}
\begin{figure}
\begin{picture}(30,6)
\thinlines
\put(0,4){\framebox(1.8,2){$a_1$}}
\put(2,4){\thicklines \framebox(1.8,2){$a_2$}}
\put(4,4){\framebox(1.8,2){$a_3$}}
\put(6,4){\framebox(1.8,2){$a_4$}}
\put(0,0){\framebox(1.8,2){$a_3$}}
\put(2,0){\framebox(1.8,2){$a_4$}}
\put(4,0){\thicklines \framebox(1.8,2){$a_2$}}
\put(6,0){\framebox(1.8,2){$a_1$}}
\put(1,4){\line(3,-1){6}}
\put(3,4){\thicklines \line(1,-1){2}}
\put(5,4){\line(-2,-1){4}}
\put(7,4){\line(-2,-1){4}}
\put(10,4){\dashbox{0.2}(1.8,2){$b_1$}}
\put(12,4){\dashbox{0.2}(1.8,2){$b_2$}}
\put(10,0){\dashbox{0.2}(1.8,2){$b_2$}}
\put(12,0){\dashbox{0.2}(1.8,2){$b_1$}}
\curl{11}{4}{13}{2}
\curl{13}{4}{11}{2}
\put(20,4){\framebox(1.8,2){$a_1$}}
\put(22,4){\dashbox{0.2}(1.8,2){$b_1$}}
\put(24,4){\dashbox{0.2}(1.8,2){$b_2$}}
\put(26,4){\framebox(1.8,2){$a_3$}}
\put(28,4){\framebox(1.8,2){$a_4$}}
\put(15,3){\makebox(4,2){resolve}}
\put(15,3){\vector(1,0){4}}
\put(20,0){\framebox(1.8,2){$a_3$}}
\put(22,0){\framebox(1.8,2){$a_4$}}
\put(24,0){\dashbox{0.2}(1.8,2){$b_2$}}
\put(26,0){\dashbox{0.2}(1.8,2){$b_1$}}
\put(28,0){\framebox(1.8,2){$a_1$}}
\put(21,4){\line(4,-1){8}}
\put(27,4){\line(-3,-1){6}}
\put(29,4){\line(-3,-1){6}}
\curl{23}{4}{27}{2}
\curl{25}{4}{25}{2}
\end{picture}
\caption{The derived permutation $\varrho$ for the resolvent \label{order-illus}}
\end{figure}
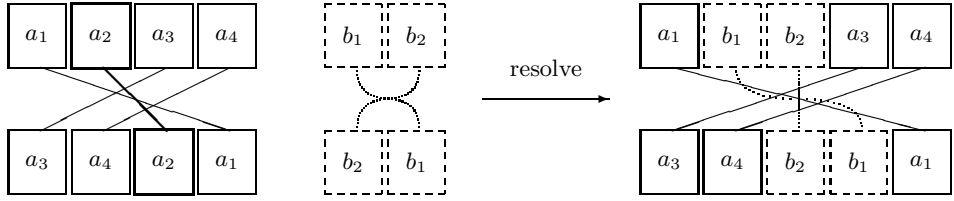

\noindent
Note that derivations of a permutation nicely moded
query and a permutation
nicely moded, input-linear program are {\em occur-check free}. This is
is a trivial consequence of Thm.~13~\cite{AL95}.
In Subsec.~\ref{weaken-subsec}, we discuss ways in which the condition
of input-linearity in Lemma~\ref{nicely-moded-persistence} can be
weakened.

\subsection{Permutation Well Typed Programs} \label{no-flounder-subsec}
In a {\em well typed} query~\cite{AL95,AP94,BLR92}, the first atom 
is correctly typed in its
input positions. Furthermore, given a well typed query
$Q,a,Q'$ and assuming LD-derivations, if $Q$ is resolved away, then $a$
becomes correctly typed in its input positions. 
We generalise this to {\em \perm well typed} (previously called
{\em properly typed}~\cite{A97}). As with the modes, we
assume that the type associated with each argument position is given.
In the examples, the types will be the natural ones that would be expected.

\begin{defi}[\perm well typed] \label{well-typed}
Let $Q =  
p_1(\vect{s}_1,\vect{t}_1),\dots,
p_n(\vect{s}_n,\vect{t}_n)$
be a query, where
$p_i(\vect{S}_i,\vect{T}_i)$
is the type of $p_i$ for each
$i\in\oneton$.
Let $\pi$ be a permutation on $\oneton$.
Then $Q$
is {\bf $\pi$-well typed} if 
for all $i \in \{1,\dots,n\}$ and $L=1$
\setcounter{equation}{0}
\begin{equation}
\models
( \bigwedge_{L \leq \pi(j) < \pi(i)}
  \vect{t}_j: \vect{T}_j ) 
\Rightarrow \vect{s}_i: \vect{S}_i. 
\label{well-typed-eq}
\end{equation}
The clause 
$p(\vect{t}_0,\vect{s}_{n+1}) \leftarrow Q$,  
where $p(\vect{T}_0,\vect{S}_{n+1})$ is the type of $p$,
is {\bf $\pi$-well typed} if (\ref{well-typed-eq}) holds 
for all $i \in \{1,\dots,n+1\}$ and $L = 0$.

A {\bf permutation well typed} 
query (clause, program)  and
a {\bf well typed} query (clause, program) 
{\bf corresponding to} a query (clause, program) are defined in analogy to 
Def.~\ref{nicely-moded}. 
\stopper \end{defi}

\begin{example} \label{well-typed-ex}
\begin{sloppypar}
Consider the program in Fig.~\ref{permute-loops-fig} with type 
$\{{\tt permute}(list,list)$,  ${\tt delete}(any,list,list)\}$. 
It  is well typed for mode 
$M_1$, and 
\perm well typed for mode 
  $M_2$,
  with the same permutations as in
Ex.~\ref{permute-loops}.
The same holds assuming type
$\{{\tt permute}(\numlist,\numlist)$, 
${\tt delete}(num,\numlist,\numlist)\}$.
\stopper 
\end{sloppypar}
\end{example}

\noindent
Permutation well-typedness is also a persistent condition.
The proof is analogous to Lemma~\ref{nicely-moded-persistence}, but using 
Lemma~23 instead of Lemma~11~\cite{AL95}.

\begin{rmlemma} \label{well-typed-persistence}
Let $Q =  a_1,\dots,a_n$ be a $\pi$-well typed query and 
$C = h \leftarrow b_1,\dots,b_m$ be a $\rho$-well typed clause 
where $\vars(Q) \cap \vars(C) = \emptyset$.
Suppose for some $k \in \oneton$, $h$ and $a_{k}$ are unifiable. 
Then the resolvent of $Q$ and $C$ with selected atom $a_k$ is 
$\varrho$-well typed, where $\varrho$ is the derived permutation  
(see Lemma~\ref{nicely-moded-persistence}).
\end{rmlemma}

\noindent
Generalising Thm.~26~\cite{AL95}, permutation well-typedness can be
used to show that derivations do not flounder~\cite{smaus-thesis}.

\subsection{Permutation Simply Typed Programs} \label{simply-typed-subsec}
We now define 
{\em \perm simply-typedness}. 
The name {\em simply typed} is a combination of 
{\em simply moded}~\cite{AL95} 
and {\em well typed}. 
In a  \perm simply typed query,
the output positions are filled with variables, and therefore they 
can always be instantiated so that all atoms
in the query are correctly typed. 

\begin{defi}[\perm simply typed] \label{simply-typed}
Let $Q =  
p_1(\vect{s}_1,\vect{t}_1),\dots,
p_n(\vect{s}_n,\vect{t}_n)$
be a query and
$\pi$ a permutation on $\oneton$.
Then $Q$
is {\bf $\pi$-simply typed} if it is 
\mbox{$\pi$-nicely} moded and $\pi$-well typed,  and
$\vect{t}_1,\dots,\vect{t}_n$ is a vector of {\em variables}.

The clause 
$p(\vect{t}_0,\vect{s}_{n+1}) \leftarrow  Q$
is {\bf $\pi$-simply typed} 
if it is 
$\pi$-nicely moded and $\pi$-well typed, 
$\vect{t}_1,\dots,\vect{t}_n$ is a vector of {\em variables}
and $\vect{t}_0$ is a vector of flat type-consistent terms that 
has a variable in each position of variable type.

A {\bf permutation simply typed} 
query (clause, program)  and
a {\bf simply typed} query (clause, program) 
{\bf corresponding to} a query (clause, program) are defined in analogy to 
Def.~\ref{nicely-moded}. 
\stopper \end{defi}

\begin{example} \label{qs-ex}
Figure~\ref{qs-fig} shows a version of the $\tt quicksort$ program.
Assume the type
 $\{ {\tt qsort}(\numlist,\numlist)$, 
 $   {\tt append}(\numlist,\numlist,\numlist)$,  
 $   {\tt leq}(num,num)$, 
 $   {\tt grt}(num,num),$
 $   {\tt part}(\numlist,num,\numlist,\numlist)\}$.
The program is \perm simply typed for mode $M_1$.
It is not \perm simply typed for mode
$M_2$, 
due to the non-variable term 
\verb@[X|Bs2]@ in an output position.
\stopper 
\end{example}

\begin{figure} 
\figrule
\begin{center}
\begin{minipage}[t]{6cm}
\small
\begin{verbatim}
:- block qsort(-,-).
qsort([],[]).
qsort([X|Xs],Ys) :-
  append(As2,[X|Bs2],Ys),
  part(Xs,X,As,Bs),
  qsort(As,As2),
  qsort(Bs,Bs2).

:- block append(-,?,-).
append([],Y,Y).
append([X|Xs],Ys,[X|Zs]) :-
  append(Xs,Ys,Zs).
\end{verbatim}
\end{minipage}
\begin{minipage}[t]{6cm}
\small
\begin{verbatim}
:- block part(?,-,?,?), 
         part(-,?,-,?), 
         part(-,?,?,-).
part([],_,[],[]).
part([X|Xs],C,[X|As],Bs):-
  leq(X,C),
  part(Xs,C,As,Bs).
part([X|Xs],C,As,[X|Bs]):-
  grt(X,C),
  part(Xs,C,As,Bs).

:- block leq(?,-), leq(-,?).
leq(A,B) :- A =< B.

:- block grt(?,-), grt(-,?).
grt(A,B) :- A > B.
\end{verbatim}
\end{minipage}
\end{center}

\vspace*{2ex}
$M_1 =\{\tt qsort(\inp,\out), 
\tt append(\inp,\inp,\out), 
\tt leq(\inp,\inp), 
\tt grt(\inp,\inp), 
\tt part(\inp,\inp,\out,\out)\}$\\
$M_2 = \{\tt qsort(\out,\inp), 
\tt append(\out,\out,\inp), 
\tt leq(\inp,\inp), 
\tt grt(\inp,\inp),
\tt part(\out,\inp,\inp,\inp)\}$ 
\caption{The {\tt quicksort} program \label{qs-fig}}
\figrule
\end{figure}

\noindent
The persistence properties stated in
Lemmas~\ref{nicely-moded-persistence} and~\ref{well-typed-persistence}
are independent of the selectability of an atom in a query.  
For \perm simply typed programs, this persistence property only holds 
if the selected atom is sufficiently
instantiated in its input arguments.
This motivates the following definition. 

\begin{defi}[bound/free] \label{bound}
Let $P$ be a \perm well typed program. An input position of a
predicate $p$ in $P$ is {\bf bound} if there is a clause head 
$p(\dots)$ in $P$ that has a non-variable term in that position.
An output position of a predicate $p$ in $P$ is {\bf bound} if 
there is an atom $p(\dots)$ in a clause body in $P$ 
that has a non-variable term in that position.
A position is {\bf free} if it is not bound.

We denote the projection of a vector of arguments $\vect{r}$ onto
its free positions as $\vect{r}^{\sf f}$,
 and onto
its bound positions as $\vect{r}^{\sf b}$.
\stopper \end{defi}

\noindent
Note that for a permutation simply typed program, there are no bound
output positions, and bound input positions must be of non-variable
type.

\begin{rmlemma} \label{simply-typed-persistence}
Let $Q =p_1(\vect{s}_1,\vect{t}_1),\dots,p_n(\vect{s}_n,\vect{t}_n)$ 
be a $\pi$-simply typed query and  
$C= p_k(\vect{v}_0,\vect{u}_{m+1}) 
\leftarrow 
q_1(\vect{u}_1,\vect{v}_1),\dots,q_m(\vect{u}_m,\vect{v}_m)$ 
a  $\rho$-simply typed, input-linear clause where 
$\vars(C) \cap \vars(Q) = \emptyset$. 
Suppose that for some $k \in \oneton$, 
$\vect{s}_k$ is non-variable in all bound input
positions\footnote{
This is similar to the assumption ``the delay declarations imply
  matching''~\cite{AL95}.}
and $\theta$ is the MGU of $p_k(\vect{s}_k,\vect{t}_k)$ and 
$p_k(\vect{v}_0,\vect{u}_{m+1})$. Then 
\begin{enumerate}
\item
there exist substitutions $\theta_1$, $\theta_2$ such that
$\theta = \theta_1\theta_2$ and
  \begin{enumerate}
  \item \label{dom1}
  $\vect{v}_0\theta_1 = \vect{s}_k$ and
  $\dom(\theta_1) \subseteq \vars(\vect{v}_0)$,
  \item \label{dom2}
  $\vect{t}_k\theta_2 = \vect{u}_{m+1} \theta_1$ and 
  $\dom(\theta_2) \subseteq \vars(\vect{t}_k)$;
  \end{enumerate}
\item \label{dom_total} 
$\dom(\theta) 
\subseteq \vars(\vect{t}_k) \cup \vars(\vect{v}_0)$;
\item  \label{orig_var}
$\dom(\theta) \cap \vars(\vect{t}_1,\dots,\vect{t}_{k-1},
\vect{v}_1,\dots,\vect{v}_m,
\vect{t}_{k+1},\dots,\vect{t}_n) =
\emptyset$;
\item \label{simply-typed-persistence-main}
the resolvent of $Q$ and $C$ with selected atom 
$p_k(\vect{s}_k,\vect{t}_k)$ is $\varrho$-simply typed, where $\varrho$
is the derived permutation
(see Lemma~\ref{nicely-moded-persistence}).
({\em Proof see Appendix})
\end{enumerate}
\end{rmlemma}

\noindent
The following corollary of 
Lemma~\ref{simply-typed-persistence} (\ref{simply-typed-persistence-main})
holds since by Def.~\ref{well-typed}, the leftmost atom in a simply
typed query is non-variable in its input positions of non-variable
type.

\begin{coro} \label{simply-typed-persistence-ld}
  Every LD-resolvent of a simply typed query $Q$ and a 
  simply typed, input-linear
clause $C$, where $\vars(C) \cap \vars(Q) = \emptyset$,
is simply typed.\footnote{This even holds
  without requiring $C$ to be
  input-linear \cite[(Smaus, 1999, Lemma~7.3)]{smaus-thesis}, 
but here we do not 
need the stronger result, and it is not a corollary of 
Lemma~\ref{simply-typed-persistence} (\ref{simply-typed-persistence-main}).}
\end{coro}

\noindent
Before studying permutation simply typed programs any further, we
now introduce a generalisation of this class.

\subsection{Permutation Robustly Typed Programs}

The program in Fig.~\ref{qs-fig} is not permutation simply typed in
mode $M_2$, due to the non-variable term
\verb@[X|Bs2]@ in an output position.
It has been acknowledged previously that it is difficult
to reason about queries where non-variable terms in output positions
are allowed, but on the other hand, there are natural programs
where this occurs~\cite{AE93}. 

We define \perm {\em robustly-typedness}, which is a carefully crafted
extension of \perm simply-typedness, allowing for non-variable but
flat terms in 
output positions. It has been designed so that a 
persistence property analogous to 
Lemmas~\ref{nicely-moded-persistence}, \ref{well-typed-persistence}
and \ref{simply-typed-persistence} holds.

\begin{defi}[\perm robustly typed] \label{robustly-typed}
Assume a permutation well 
typed program $P$ where the bound positions are
of non-variable type. Let $Q =  
p_1(\vect{s}_1,\vect{t}_1),\dots,p_n(\vect{s}_n,\vect{t}_n)$
be a query (using predicates from $P$) and
$\pi$ a permutation on $\oneton$.
Then $Q$
is {\bf $\pi$-robustly typed} if it is 
$\pi$-nicely moded and $\pi$-well typed, 
$\vect{t}^{\sf f}_1,\dots,\vect{t}^{\sf f}_n$ is a vector of variables, 
and  
$\vect{t}^{\sf b}_1,\dots,\vect{t}^{\sf b}_n$ is a vector of flat
type-consistent terms.

The clause 
$p(\vect{t}_0,\vect{s}_{n+1}) \leftarrow
Q$
is {\bf $\pi$-robustly typed} if it is 
$\pi$-nicely moded; $\pi$-well typed; 
\begin{enumerate}
\item  
$\vect{t}^{\sf f}_0,  \dots, \vect{t}^{\sf f}_n$ 
is a vector of variables, and 
$\vect{t}_0^{\sf b},\dots,\vect{t}^{\sf b}_n$ is a vector of flat
type-consistent terms; and
\item\label{var-in-b-pos} 
if a position in $\vect{s}^{\sf b}_{n+1}$ of type $\tau$ is filled
  with a variable $x$, then $x$ also fills a position of type $\tau$
  in $\vect{t}^{\sf b}_0,\dots,\vect{t}^{\sf b}_n$.
\end{enumerate} 
A {\bf permutation robustly typed} 
query (clause, program) and
a {\bf robustly typed} query (clause, program) 
{\bf corresponding to} a query (clause, program) 
are defined in analogy to 
Def.~\ref{nicely-moded}.
\stopper \end{defi}

\noindent
Note that any permutation simply typed program is permutation robustly
typed in which all output positions are free.

\begin{example} \label{qs-robustly-ex}
Recall the program in Fig.~\ref{qs-fig}.
It is
\perm robustly typed in mode $M_2$, and
the second position of {\tt append} is the only bound output position.
Note in particular that Condition~\ref{var-in-b-pos} of
Def.~\ref{robustly-typed} is met for the recursive clause of
{\tt append}: the variable {\tt Ys} fills an output position of the head
and also an output position of the body.
Moreover, the program is trivially \perm robustly typed in mode $M_1$.
\end{example}

\begin{figure} 
\figrule
\begin{center}
\begin{minipage}[t]{7cm}
\small
\begin{verbatim}
:- block treeList(-,-).
treeList(leaf,[]).
treeList(node(L,Label,R),List) :-
  append(LList,[Label|RList],List),
  treeList(L,LList),
  treeList(R,RList).
\end{verbatim}
\end{minipage}
\begin{minipage}[t]{5cm}
\small
\begin{verbatim}
:- block append(-,?,-).
append([],Y,Y).
append([X|Xs],Ys,[X|Zs]) :-
  append(Xs,Ys,Zs).
\end{verbatim}
\end{minipage}
\end{center}

\vspace*{2ex}
$M_1 = \{ \tt treeList(\inp,\out), append(\inp,\inp,\out) \}$\\
$M_2 = \{ \tt treeList(\out,\inp), append(\out,\out,\inp) \}$\\
\caption{Converting trees to lists or vice versa\label{ttl-fig}}
\figrule
\end{figure}

\begin{example}\label{ttl-robustly-ex}
The program in Fig.~\ref{ttl-fig} converts binary trees into lists and vice
versa. Assuming type
$\{{\tt treeList}(tree,list)$, ${\tt append}(list,list,list)\}$,
the program is \perm robustly typed in mode $M_2$, and
the second position of {\tt append} is the only bound output position.
It is also \perm robustly typed in mode
$M_1$, where all output
positions are free.
\stopper \end{example}

\noindent
The following lemma shows a persistence property of \perm 
robustly-typedness, and shows, furthermore, that a
derivation step cannot instantiate the input arguments of the selected
atom. 

\begin{rmlemma} \label{robustly-typed-persistence}
Let $Q =  
p_1(\vect{s}_1,\vect{t}_1),\dots,p_n(\vect{s}_n,\vect{t}_n)$
a $\pi$-robustly typed query and 
$C = 
p_k(\vect{v}_0,\vect{u}_{m+1}) \leftarrow
q_1(\vect{u}_1,\vect{v}_1),\dots,q_m(\vect{u}_m,\vect{v}_m)$
a $\rho$-robustly typed, input-linear clause where
$\vars(Q) \cap \vars(C) = \emptyset$. 
Suppose that for some $k \in \oneton$, 
$p_k(\vect{s}_k,\vect{t}_k)$ is non-variable in all bound input 
positions
and $\theta$ is the MGU of $p_k(\vect{s}_k,\vect{t}_k)$ and 
$p_k(\vect{v}_0,\vect{u}_{m+1})$. 

Then the resolvent of $Q$ and $C$ with selected atom $p_k(\vect{s}_k,\vect{t}_k)$ is 
$\varrho$-robustly typed, where $\varrho$ is the derived permutation
(see  Lemma~\ref{nicely-moded-persistence}). Moreover
$\dom(\theta) \cap \vars(\vect{s}_k) = \emptyset$.
({\em Proof see Appendix})
\end{rmlemma}

\noindent
We now define programs where the \block\ 
declarations fulfill a natural minimum 
(\ref{min-sel}) and maximum (\ref{max-sel})
requirement. The minimum requirement states that selected atoms
must fulfill the assumption of
Lemmas~\ref{simply-typed-persistence}
and~\ref{robustly-typed-persistence}. The maximum requirement is needed
in Subsec.~\ref{well-fed-subsec}. In other words, we define programs
where the ``static'' concept of modes and the ``dynamic'' concept 
of \block\ declarations correspond in the natural way.

\begin{defi}[input selectability] \label{selectable2}
Let $P$ be a \perm robustly typed 
program. $P$ has {\bf input selectability}
if an atom using a predicate in $P$ that has variables in all
free output positions is selectable in $P$ 
\begin{enumerate}
\item \label{min-sel}
only if it is non-variable in all bound input 
positions; and
\item \label{max-sel}
if it is non-variable in all input positions of non-variable type.
\end{enumerate}
\stopper 
\end{defi}

\noindent
Note that the above definition is aimed at atoms in permutation
robustly typed queries, since these atoms have variables in all free output
positions. 

\begin{example} \label{sel-ex}
Consider $\tt append(\out,\out,\inp)$ where the second position is the
only bound output position, as used in the programs in
Fig.~\ref{qs-fig} in mode $M_2$ and Fig.~\ref{ttl-fig} in mode $M_2$.
The program for $\tt append$ has input selectability.

Now consider $\tt append(\inp,\inp,\out)$ where the output position is
free, as used in the programs in
Fig.~\ref{qs-fig} in mode $M_1$ and Fig.~\ref{ttl-fig} in mode $M_1$.
The program for $\tt append$ has input selectability.
Note that the \block\ declaration for {\tt append} is the one that is 
usually given~\cite{goedel,L93,MT99}.
\stopper
\end{example}

\noindent
The following is a corollary of
Lemma~\ref{robustly-typed-persistence} needed to prove Lemma~\ref{needs-feeding}.

\begin{coro} \label{consumer-affects-not}
Let $P$ be a \perm robustly typed, input-linear program with input selectability, 
$Q = a_1,\dots,a_n$ a $\pi$-robustly typed query and 
$i,j \in \oneton$ such that $\pi(i) < \pi(j)$.
Let 
\[
Q \steps \;
(b_1,\dots,b_m) 
\step \;
(b_1,\dots,b_{l-1},B,b_{l+1},\dots,b_m) \theta
\]
 be a delay-respecting derivation and $k \in \onetom$,  
such that  $b_k$ is a descendant of $a_i$
and $b_l$ is a descendant of $a_j$. 
Then $\dom(\theta) \cap \vars(b_k) = \emptyset$.
\end{coro} 

\begin{proof}
Suppose that $b_1,\dots,b_m$ is $\rho$-robustly typed.
By Cor.~\ref{mode-order} we have $\rho(k) < \rho(l)$.
Suppose
$b_l=p_l(\vect{s}_l,\vect{t}_l)$.

Since $\theta$ is obtained by unifying $b_l$ with a head of a clause $C$,
and 
$\vars(C) \cap \vars(b_1,\dots,b_m) = \emptyset$, it follows that
$\dom(\theta) \cap \vars(b_1,\dots,b_m) \subseteq \vars(b_l)$.
By Lemma~\ref{robustly-typed-persistence}, 
$\dom(\theta) \cap \vars(\vect{s}_l) = \emptyset$.
Since $b_1,\dots,b_m$ is $\rho$-nicely moded,  
$\vars(b_k) \cap \vars(\vect{t}_l) = \emptyset$ 
 and so 
$\dom(\theta) \cap  
\vars(b_k) = \emptyset$.
 \end{proof}

\noindent
Intuitively, the corollary says that if $\pi(i) < \pi(j)$, then no 
$a_j$-step will ever instantiate a descendent of $a_i$.

We conclude the section with a statement about permutation {\em
simply} typed programs, which we could not present earlier since it
relies on the definition of input selectability. 
It says that in a derivation for a \perm simply typed
program and query, it can be assumed without loss of generality that
the output positions in each query are filled with variables that
occur in the initial query or in some clause body used in the
derivation. 

\begin{coro} \label{star-dagger}
Let $P$ be a \perm simply typed program with input selectability and 
$Q_0$ be a \perm simply typed query.
Let $\theta_0 = \emptyset$ and
$\xi = 
\langle Q_0, \theta_0 \rangle; 
\langle Q_1, \theta_1 \rangle; 
\dots$ be a delay-respecting derivation of $P \cup \{ Q_0 \}$.
Then for all $i \geq 0$,
if $x$ is a variable in an output position in
$Q_i$, then $x \theta_i = x$.
\end{coro}

\begin{proof}
The proof is by induction on the position $i$ in the derivation.
The base case $i =0$ is trivial since $\theta_0 = \emptyset$.
Now suppose the result holds for some $i$ and $Q_{i+1}$ exists.
By Lemma~\ref{simply-typed-persistence} (\ref{simply-typed-persistence-main}), 
$Q_{i} \theta_{i}$ is \perm simply typed.
Thus the result follows for $i+1$ by 
Lemma~\ref{simply-typed-persistence}~(\ref{orig_var}).
\end{proof}

\section{Termination and Speculative Bindings} \label{termination-a}

Like most approaches to the termination problem~\cite{DD94}, 
we are interested in ensuring that {\em all}
derivations of a query are finite. Therefore the clause order in a
program is irrelevant.
Furthermore, we do not prove termination as such, but rather 
reduce the problem of proving termination for a program and query
with left-based derivations to that with
LD-derivations.

In this section, we present two complementing methods of showing
termination. These are explained in the following example.

\begin{figure}
\figrule
\begin{center}
\begin{minipage}[t]{6cm}
\small
\begin{verbatim}
:- block permute(-,-).
permute([],[]).
permute([U|X],Y) :- 
  delete(U,Y,Z),
  permute(X,Z).
\end{verbatim}
\end{minipage}
\begin{minipage}[t]{6cm}
\small
\begin{verbatim}
:- block delete(?,-,-).
delete(X,[X|Z],Z).
delete(X,[U|Y],[U|Z]) :- 
  delete(X,Y,Z).
\end{verbatim}
\end{minipage}
\end{center}

\vspace*{2ex}
$M_1 = \{ {\tt permute(\inp,\out)},
{\tt delete(\inp,\out,\inp)}\}$\\
$M_2 = \{ {\tt permute(\out,\inp)}, 
{\tt delete(\out,\inp,\out)}\}$
\caption{Putting recursive calls last in the {\tt permute} program \label{permute-ok-fig}}
\figrule
\end{figure}

\begin{example} \label{permute-ok} 
Assuming left-based derivations, 
the program in Fig.~\ref{permute-loops-fig} loops 
for the query 
{\tt permute(V,[1])} (hence, in mode 
$M_2$)
because {\tt delete} produces 
a {\em speculative output binding}~\cite{N92}: The third argument of 
{\tt delete} is bound 
before  it is known that this binding will never have to be undone.
Termination in modes $M_1$ and $M_2$
 can be ensured by swapping the atoms in the second clause, as shown
in Fig.~\ref{permute-ok-fig}.
This technique has been described as {\em putting recursive calls
  last}~\cite{N92}.  To explain why the program terminates, we have 
to apply a
different reasoning for the different modes.  

In mode $M_2$, the atom that produces the
speculative output occurs {\em textually before} the atom that
consumes it. This means that the consumer waits until the producer has
completed (undone the speculative binding).  The program does not {\em
  use} speculative bindings.  In mode $M_1$, the program does not 
{\em make} speculative bindings.
  
Note that termination for this example depends on {\em left-based}
derivations, and thus any method that abstracts from the selection
rule must fail.  
\stopper \end{example}

\noindent
The methods presented in this section can be used to prove that the
programs 
in Figs.~\ref{ttl-fig}, \ref{permute-ok-fig}, 
\ref{specific-delete-fig}, and \ref{only-nonspec-works-fig}
terminate, but they do not work for the programs in Figs.~\ref{qs-fig} 
and~\ref{nqueens-fig}. 
They formalise previous heuristics~\cite{N85,N92} and
 rely on conditions that are easy to check.

\subsection{Termination by not Using Speculative Bindings} \label{not-using}
In LD-derivations, speculative
bindings are never used~\cite{N92}. A left-based der\-i\-va\-tion {\em is} 
an LD-derivation, provided the leftmost
atom in each query is always selectable. 
Hence by Lemma~\ref{well-typed-persistence}, 
we have the following proposition.

\begin{propo} \label{left=ld}
Let $Q$ be a well typed query and $P$ a well typed program
such that an atom is selectable in $P$
whenever its input positions of
non-variable type are non-variable. 
Then every left-based derivation of $P \cup \{ Q \}$ is an LD-derivation.
\end{propo} 

\noindent
We now give two examples of programs where by
Proposition~\ref{left=ld}, we can use any
method for LD-derivations to show termination
for any well typed query. 
Note that the method of Sec.~\ref{termination-b} is not applicable 
for the program in Ex.~\ref{most-specific}
(because it is is not \perm robustly typed).

\begin{example} \label{left=ld-ex}
Consider the program in Fig.~\ref{permute-ok-fig} with mode 
$M_2$
and either of the types given in Ex.~\ref{well-typed-ex}.  This
program is well-typed. 
\stopper \end{example}

\begin{figure}
\figrule
\begin{center}
\small
\begin{verbatim}
:- block delete(?,-,-).
delete(X,[X|Z],Z).
delete(X,[U|[H|T]],[U|Z]) :- delete(X,[H|T],Z).
\end{verbatim}
\end{center}
\caption{Most specific version of $\tt delete(\out,\inp,\out)$\label{specific-delete-fig}}
\figrule
\end{figure}

\begin{example} \label{most-specific}
Consider the version of 
$\tt delete(\out,\inp,\out)$ given in Fig.~\ref{specific-delete-fig}.
Assuming either 
of the types given in Ex.~\ref{well-typed-ex}, 
this program is well typed. 
\stopper 
\end{example}

\noindent
Regarding this subsection, one may wonder: what is the point in
considering derivations for programs with \block\ declarations where
in effect we show that those \block\ declarations are redundant, that
is, the program is executed left-to-right? However, one has to
bear in mind that a program might also be used in another mode, and
therefore, the \block\ declarations may be necessary. 

\subsection{Termination by not Making Speculative Bindings}\label{nonspec-subsec}
Some programs and queries have the nice property that there cannot be
any failing derivations. Bossi and Cocco~\shortcite{BC98} have identified
a class of programs called
{\em noFD} having this property.  {\em
Non-speculative} programs are similar, but there are two differences:
the definition of noFD programs only allows for {\em LD}-derivations,
but on the other hand, the definition of non-speculative programs
requires that the clause heads are input-linear.

\begin{defi}[non-speculative] \label{nonspec}
A program $P$ is {\bf non-speculative} if it is \perm simply typed and
input-linear, and every simply typed atom using a predicate in $P$ is
unifiable with some clause head in $P$.  \stopper \end{defi}

\begin{example}
Both versions of the {\tt permute} program 
(Figs.~\ref{permute-loops-fig} and \ref{permute-ok-fig}), with 
either type 
given in Ex.~\ref{well-typed-ex}, 
are non-speculative in mode  
$M_1$.
Every simply typed atom is unifiable with at least one clause
head.
Both versions are 
{\em not} non-speculative in mode
$M_2$, because 
\verb@delete(A,[],B)@
is not unifiable with any clause head.
\stopper 
\end{example}

\begin{example} 
The program in Fig.~\ref{ttl-fig} is non-speculative in mode 
$M_1$. However it is not non-speculative in mode 
$M_2$
because it is not \perm simply
typed, due to the non-variable term 
\verb@[Label|List]@ in an output position. 
\stopper \end{example}

\noindent
A delay-respecting derivation for a non-speculative program $P$ with
input selectability and a
\perm simply typed query cannot fail.\footnote{
It can also not {\em flounder}~\cite{smaus-thesis}.}
However it could still be infinite. 
The following theorem says that this
can only happen if the simply typed program corresponding to $P$ 
has an infinite {\em LD}-derivation
for this query. 

\begin{rmtheorem} \label{loop-relation}
Let $P$ be a non-speculative program with input selectability 
and $P'$ a simply typed program
corresponding to $P$.
Let $ Q$ be a
\perm simply typed query and $ Q'$ a simply typed 
query corresponding to $Q$. 
If there is an infinite delay-respecting derivation of
$P \cup \{ Q \}$, 
then there is  an infinite LD-derivation of
$P' \cup \{ Q' \}$.
({\em Proof see Appendix})
\end{rmtheorem}

\noindent
Theorem~\ref{loop-relation} says that for non-speculative programs,
the atom order in clause bodies is irrelevant for termination.

Note that any program that uses {\em tests} cannot be non-speculative.
In Fig.~\ref{qs-fig}, assuming mode $M_1$, the atoms \verb@leq(X,C)@ and
\verb@grt(X,C)@ are tests. These tests are {\em exhaustive}, that is, at
least one of them succeeds~\cite{BC98}. This suggests a generalisation
of non-speculative programs~\cite{PR99} (see Sec.~\ref{related}). 

We now give an example of a program for which 
termination can be shown using Thm.~\ref{loop-relation} but 
not using the method of Sec.~\ref{termination-b} 
(see also Ex.~\ref{only-nonspec-works-cont}).

\begin{figure}
\figrule
\begin{center}
\begin{minipage}[t]{6cm}
\small
\begin{verbatim}
:- block is_list(-).
is_list([]).
is_list([X|Xs]):-
  is_list(Ys),
  equal_list(Xs,Ys).
\end{verbatim}
\end{minipage}
\begin{minipage}[t]{6cm}
\small
\begin{verbatim}
:- block equal_list(-,?).
equal_list([],[]).
equal_list([X|Xs],[X|Ys]):-
  equal_list(Xs,Ys).
\end{verbatim}
\end{minipage}
\end{center}
\caption{The $\tt is\_list$ program \label{only-nonspec-works-fig}} 
\figrule
\end{figure}

\begin{example}\label{only-nonspec-works}
Consider the program in Fig.~\ref{only-nonspec-works-fig}, 
where the mode is 
$\{\tt
is\_list(\inp),$ 
$\tt equal\_list(\inp,\out)
\}$
and the type is
$\{ 
{\tt is\_list}(list),$ 
${\tt equal\_list}(list,list)
\}$.
The program is permutation simply typed (the second clause is 
$\langle2,1\rangle$-simply typed) and non-speculative, and 
all LD-derivations for the
corresponding simply typed program terminate.
Therefore all delay-respecting derivations of a
permutation simply typed query and this program terminate.
\stopper
\end{example}

\section{Termination and Insufficient Input} \label{termination-b}

We will now present an alternative method for showing 
termination that overcomes some of the limitations of the methods
presented in the previous section. In particular, the methods 
can be used for the programs in Figs.~\ref{qs-fig} 
and~\ref{nqueens-fig} as well 
as Figs.~\ref{ttl-fig} and~\ref{permute-ok-fig}. In practice, we expect the method
presented here to be more useful, although, as
Figs.~\ref{specific-delete-fig} and~\ref{only-nonspec-works-fig}
show, it does not subsume the method of the
previous section.

As explained in Ex.~\ref{permute-ok}, termination of
$\tt permute(\out,\inp)$ can be ensured by applying the heuristic of putting
recursive calls last~\cite{N92}. The following example however shows
that even this version of $\tt permute(\out,\inp)$ can cause a loop
depending on how it is called within some other program.

\begin{figure}
\figrule
\begin{minipage}[t]{5cm}
\small
\begin{verbatim}
:- block nqueens(-,?).
nqueens(N,Sol) :-
  sequence(N,Seq),
  safe(Sol),        
  permute(Sol,Seq).

:- block sequence(-,?).
sequence(0,[]).
sequence(N,[N|Seq]):-
  0 < N,
  N1 is N-1,
  sequence(N1,Seq).

:- block safe(-).
safe([]).
safe([N|Ns]) :-
  safe_aux(Ns,1,N),
  safe(Ns).

\end{verbatim}
\end{minipage}
\begin{minipage}[t]{7cm}
\small
\begin{verbatim}
:- block safe_aux(-,?,?), safe_aux(?,-,?), 
                          safe_aux(?,?,-).
safe_aux([],_,_).
safe_aux([M|Ms],Dist,N) :-
  no_diag(N,M,Dist),
  Dist2 is Dist+1,
  safe_aux(Ms,Dist2,N).

:- block no_diag(-,?,?), no_diag(?,-,?).
no_diag(N,M,Dist) :-
  Dist =\= N-M,
  Dist =\= M-N.

:- block permute(-,-).
permute([],[]).
permute([U|X],Y) :- 
  delete(U,Y,Z),
  permute(X,Z).

:- block delete(?,-,-).
delete(X,[X|Z],Z).
delete(X,[U|Y],[U|Z]) :- 
  delete(X,Y,Z).
\end{verbatim}
\end{minipage}

\vspace*{2ex}
$
\begin{array}{rcl}
M_1 & = & 
\{ \tt nqueens(\inp,\out), 
\tt sequence(\inp,\out),
\tt safe(\inp),
\tt permute(\out,\inp), 
\tt \verb@<@(\inp,\inp),\\
& & \hspace{0.4em}
\tt is(\out,\inp), 
\tt safe\_aux(\inp,\inp,\inp),
\tt no\_diag(\inp,\inp,\inp),
\tt \verb@=\=@ (\inp,\inp) \}\\
M_2 & = & 
\{ \tt nqueens(\out,\inp), 
sequence(\out,\inp),
permute(\inp,\out),
is(\out,\inp), \dots \}\\
T & = & 
\{ 
{\tt nqueens}(int,\intlist), 
{\tt sequence}(int,\intlist), 
{\tt safe}(\intlist), 
{\tt permute}(\intlist,\intlist),\\
& & \hspace{0.4em} 
\verb@<@(int,int),
{\tt is}(int,int), 
{\tt safe\_aux}(\intlist,int,int),
{\tt no\_diag}(int,int,int),\\
& & \hspace{0.4em} 
\verb@=\=@ (int,int)
\}
\end{array}
$

\caption{A program for $n$-queens \label{nqueens-fig}}
\figrule
\end{figure}

\begin{example} \label{nqueens-ex}
Figure~\ref{nqueens-fig} shows a program
for  the $n$-queens problem, which uses \block\
declarations to
implement the test-and-generate 
paradigm. 
With the mode $M_1$ and the type $T$, the first 
clause is  $\langle 1,3,2 \rangle$-nicely moded and 
$\langle 1,3,2 \rangle$-well typed. 
Moreover, all left-based derivations for 
the query  \verb@nqueens(4,Sol)@ terminate.

\begin{sloppypar}
However, if in the first clause, 
the atom order is changed by moving 
\verb@sequence(N,Seq)@ 
to the end, then 
\verb@nqueens(4,Sol)@ loops. This is because resolving
\verb@sequence(4,Seq)@ with the second clause for \verb@sequence@ 
makes a 
binding 
(which is {\em not} speculative)
that triggers the call \verb@permute(Sol,[4|T])@. 
This call results in a loop.
Note that \verb@[4|T]@, although non-variable, 
is insufficiently instantiated for
\verb@permute(Sol,[4|T])@ to be correctly typed in its
input position: 
\verb@permute@ is called with {\em insufficient input}.
\stopper 
\end{sloppypar}
\end{example}

\noindent
To ensure termination, each atom that may loop when called with
insufficient input should be placed sufficiently late; all producers
of input for that atom must occur textually earlier. This assumes
left-based derivations. Note
that this explains in particular why in the recursive clause for 
$\tt permute$, the recursive call should be placed last, and hence we are
effectively refining the heuristic proposed by Naish~\shortcite{N92}.
Note also that in nicely and well moded programs, {\em all} atoms are
placed sufficiently late in this sense.

In the next subsection, we identify
the {\em robust} predicates, which are predicates for which all
delay-respecting derivations are finite.
In Subsec.~\ref{well-fed-subsec},
we prove termination for programs where the atoms
using non-robust predicates are selected ``sufficiently late''.

\subsection{Robust Predicates}
 In this subsection, 
 derivations are {\em not} required to be left-based. 
The statements hold for arbitrary delay-respecting
 derivations, and thus the textual position of an atom in a query is
 irrelevant.  
Therefore we can, for just this subsection, assume that the programs and
 queries are
 robustly typed (rather than just {\em permutation} robustly typed).
This simplifies the notation. In Subsec.~\ref{well-fed-subsec}, we go
 back to allowing for arbitrary permutations.

\begin{defi}[robust] \label{robust}
Let $P$ be a robustly typed, input-linear program with input
selectability. A predicate $p$ in $P$ is {\bf robust} 
if, for each robustly typed query
$ p(\vect{s},\vect{t})$, 
any delay-respecting derivation of $P\cup\{p(\vect{s},\vect{t})\}$ 
is finite. An atom is {\bf robust}
if its predicate is.
\stopper \end{defi}

\noindent
By definition, a delay-respecting derivation for a query consisting of 
{\em one} robust atom terminates. We will see shortly 
however that this extends to queries of {\em arbitrary} length.
To prove this, we first need the following simple lemma.

\begin{rmlemma}\label{sigma-exists}
Let $Q =  
p_1(\vect{s}_1,\vect{t}_1),\dots,p_n(\vect{s}_n,\vect{t}_n)$
be a robustly typed query. Then there exists a substitution $\sigma$
such that 
$\dom(\sigma) = \vars(\vect{t}_1,\dots,\vect{t}_{n-1})$, and
$p_n(\vect{s}_n,\vect{t}_n)\sigma$ is robustly typed.
\end{rmlemma}

\begin{proof}
Since $Q$ is robustly typed and types are closed under instantiation,
there exists a substitution $\sigma$ such that 
$\dom(\sigma) = \vars(\vect{t}_1,\dots,\vect{t}_{n-1})$, 
$\ran(\sigma) = \emptyset$, and
$(\vect{t}_1,\dots,\vect{t}_{n-1})\sigma$ is correctly
typed.

Since $Q$ is nicely moded, 
$\dom(\sigma) \cap \vars(\vect{t}_n) = \emptyset$.
Since $\ran(\sigma) = \emptyset$, it follows that 
$\vars(\vect{s}_n\sigma) \cap \vars(\vect{t}_n\sigma) = \emptyset$ and
hence
$p_n(\vect{s}_n,\vect{t}_n)\sigma$  is nicely moded. 

Since $Q$ is well typed, it follows by 
Def.~\ref{well-typed} that 
$p_n(\vect{s}_n,\vect{t}_n)\sigma$ is well typed. 

Therefore, as $Q$ is robustly typed and 
$\vect{t}_n\sigma = \vect{t}_n$, it follows that 
$p_n(\vect{s}_n,\vect{t}_n)\sigma$ is robustly typed.
\end{proof}

\noindent
The following lemma says that a robust atom cannot proceed
indefinitely unless it is repeatedly ``fed'' by some other atom. 

\begin{rmlemma} \label{needs-feeding}
Let $P$ be a robustly typed, input-linear program with 
input selectability and
 $F,b,H$ a robustly typed query where 
$b$ is a robust atom.  A delay-respecting derivation of 
$P \cup \{F,b,H\}$ can
have infinitely many $b$-steps only if it has infinitely many
$a$-steps, for some $a \in F$. 
({\em Proof see Appendix})
\end{rmlemma}

\noindent
The following lemma is a consequence and states that the robust
atoms in a query on their own cannot produce an infinite derivation.

\begin{rmlemma} \label{robust=finite}
Let $P$ be a robustly typed, input-linear program with input 
selectability
and 
$Q$ a robustly typed query. 
A delay-respecting derivation of $P \cup \{Q\}$ can be infinite only if
there
are infinitely many steps where a non-robust atom is resolved.
\end{rmlemma}

\begin{proof}
Let $\xi$ be an infinite delay-respecting derivation of $P \cup
\{Q\}$. Assume, for the purpose of deriving a contradiction, that
$\xi$ contains only finitely many steps where a non-robust atom is
resolved.  Then there exists an infinite suffix $\tilde{\xi}$ of $\xi$
containing no steps where a non-robust atom is resolved.  Consider the
first query $\tilde{Q}$ of $\tilde{\xi}$.  Then there is at least one
atom in $\tilde{Q}$ that has infinitely many descendants. Let
$\tilde{a}$ be the leftmost of these atoms.  Then as $\tilde{a}$ is
robust, we have a contradiction to Lemma~\ref{needs-feeding}.  
\end{proof}

\noindent
Approaches to termination usually rely on
measuring the size of the {\em input} in a query.  We agree with
Etalle et al.~\shortcite{EBC99} that it is reasonable to make this
dependency explicit. This gives rise to the notion of {\em moded level
mapping}, which is an instance of {\em level mapping} introduced by
Bezem~\shortcite{B93} and Cavedon~\shortcite{C89}. Since we use
well typed programs instead of well moded ones~\cite{EBC99}, 
we have to generalise the concept further.

In the following definition, 
$\iatoms$ denotes the set of atoms using
predicates occurring in $P$, that are correctly typed in their input
positions. 

\begin{defi}[moded typed level mapping] \label{level-mapping}
Let $P$ be a program. 
A function $|.|: \iatoms \rightarrow \nat$
is a
{\bf moded typed level mapping} if for each
$p(\vect{s},\vect{t})\in \iatoms$

\begin{itemize}
\item 
for any $\vect{u}$, we have
$|p(\vect{s},\vect{t})| = |p(\vect{s},\vect{u})|$.
\item
for any substitution $\theta$, 
$|p(\vect{s},\vect{t})| = |p(\vect{s}\theta,\vect{t})|$.
\end{itemize}
For $a \in \iatoms$, $|a|$ is the {\bf level} of $a$.  
\stopper
\end{defi}

\noindent
Thus the level of an atom in $\iatoms$ only depends on the terms in
the input positions. Moreover, all {\em instances} of an atom in 
$\iatoms$ have the same level. Here our concept differs from moded
level mappings. Also, our concept is defined for atoms in $\iatoms$
that are not necessarily ground, but this difference only concerns the
presentation.

Since we only consider moded typed level mappings, we will
simply call them {\bf level mappings}.

The following standard concept is widely used in the termination 
literature~\cite{A97}.

\begin{defi}[depends on] \label{level}
Let $p, q$ be predicates in a program $P$. Then {\bf $p$ refers to $q$} if there
is a clause in $P$ with $p$ in its head and $q$ in its body, and 
{\bf $p$ depends on $q$} (written $p \sqsupseteq q$) if
$(p,q)$ is in the reflexive, transitive closure of {\em refers to}. 
We write $p \sqsupset q$ if $p \sqsupseteq q$ and 
$q \not\sqsupseteq p$, 
and $p \approx q$ if $p \sqsupseteq q$ and $q \sqsupseteq p$.

Abusing notation, we shall also use the above symbols for {\em atoms},
where $p(\vect{s},\vect{t}) \sqsupseteq q(\vect{u},\vect{v})$ 
stands for $p \sqsupseteq
q$, and likewise for $\sqsupset$ and $\approx$. Furthermore, we denote
the equivalence class of a predicate $p$ with respect to $\approx$ as 
$[p]_\approx$.
\stopper \end{defi}

\noindent
The following concept is used to show robustness.

\begin{defi}[well-recurrent] \label{well-recurrent}
Let $P$ be a program and $|.|$ a level mapping.
A clause $C = h \falls B$ is 
{\bf well-recurrent (with respect to $|.|$)} if, 
for every $a$ in $B$ such that $a \approx h$, and  
every substitution $\theta$ such that 
$a \theta, h \theta \in \iatoms$,
we have $|h \theta| > |a \theta|$. 

A program (set of clauses) is {\bf well-recurrent with respect to
$|.|$} if each clause is well-recurrent with respect to
$|.|$.
\stopper \end{defi}

\noindent
Well-recurrence 
resembles {\em well-acceptability}~\cite{EBC99} in
that only for atoms $a \approx h$ there has to be a decrease, and that
it assumes moded level mappings. It differs from well-acceptability,
but also from {\em delay-recurrence}~\cite{MT99}, in that it does
not refer to a model of the program.

To show that a predicate $p$ is robust, we assume that all predicates
$q$ with $p \sqsupset q$ have already been shown to be robust.

\setcounter{equation}{0}
\begin{rmlemma} \label{terminates}
Let $P$ be a robustly typed, input-linear program with input 
selectability and~$p$ a predicate in $P$. 
Suppose all predicates $q$ with $p \sqsupset q$ are 
robust, and all clauses defining predicates 
$q \in [p]_\approx$ are well-recurrent with respect to some level
mapping $|.|$. 
Then $p$ is robust. 
({\em Proof see Appendix})
\end{rmlemma}

\begin{example}
We demonstrate for the program in Fig.~\ref{nqueens-fig}, 
with mode $M_1$ and type $T$,
how Lemma~\ref{terminates} is used.\footnote{
We assume that the built-ins used here meet the conditions of
Def.~\ref{selectable2}. We will see in
Subsec.~\ref{constant-drop-block-subsec} why this is a safe assumption.}
Given  that the built-in
\verb@=\=@ terminates, it follows that 
$ {\tt no\_diag}$ is robust.
With the list length of  the first argument of $\tt safe\_aux$ as level mapping,
the clauses defining $\tt safe\_aux$ are
well-recurrent so that $\tt safe\_aux$ is robust.
In a similar way, we can show that $\tt safe$ is robust.
\stopper \end{example}

\begin{example}\label{only-nonspec-works-cont}
Consider again Ex.~\ref{only-nonspec-works}. 
We conjecture that $\tt is\_list$ is robust, 
but Lemma~\ref{terminates} cannot show this.
While Ex.~\ref{only-nonspec-works} is contrived, 
it suggests that the method
of Subsec.~\ref{nonspec-subsec} might be useful whenever 
Lemma~\ref{terminates} fails to prove that a predicate is
robust. On the other hand, one could envisage to improve the method
for showing robustness, for example by exploiting information given by
a {\em model} of the program~\cite{EBC99}. 
\stopper \end{example}

\subsection{Well Fed Programs} \label{well-fed-subsec}
As seen in Ex.~\ref{nqueens-ex}, there are predicates for which
requiring delay-respecting derivations is not sufficient for
termination.  In general, the selection rule must be taken into
account.  We assume left-based derivations.  Consequently we now give
up the assumption, made to simplify the notation, that the clauses and
query are robustly typed, rather than just {\em permutation} robustly
typed.  All statements from the previous subsection generalise to
permutation robustly typed in the obvious way.

A {\em safe} position in a query is a position that is ``sufficiently
late''. 

\begin{defi}[safe position] \label{safe-position}
For a permutation $\pi$, 
$i$ is a called {\bf safe position for} $\pi$ if 
for all $j$, $\pi(j) < \pi(i)$ implies $j < i$.
\stopper \end{defi}

\noindent
Whenever we simply speak of an {\bf atom in a safe position}, we mean
that this atom occurs in a $\pi$-robustly typed query $Q$ in the 
$i$th position and $i$ is a safe position for $\pi$, where $Q$ and
$\pi$ are clear from the context.

The next lemma says that in a left-based derivation,  
atoms whose ancestors are all in safe positions 
can never be waiting (see Def.~\ref{left-based-def}).

\begin{rmlemma} \label{safe=notwaiting}
Let $P$ be a permutation robustly typed program with input
selectability, $Q_0$ a permutation robustly typed query and 
$\xi= Q_0\steps Q_i \dots$ a left-based derivation of 
$P \cup \{Q_0\}$.  Then no atom in $Q_i$ for which all ancestors
are in safe positions is waiting.
\end{rmlemma}

\begin{proof}
Suppose $Q_i=a_1,\dots,a_n$ is $\pi_i$-robustly typed 
(note that $\pi_i$ exists by
Lemma~\ref{robustly-typed-persistence}).  Let $a_k$ be an
atom in $Q_i$ with all its ancestors in safe positions. By
Def.~\ref{well-typed}, $a_{{\pi_i}^{-1}(1)}$ is correctly typed in its
input positions and hence selectable.  Moreover, since $k$ is a safe
position, ${{\pi_i}^{-1}(1)} \leq k$.  It follows that if the
{\em proper ancestors} 
of $a_k$ are not waiting, then $a_k$ is 
not waiting.

The result follows by induction on $i$.  When $i=0$, $a_k$ has no
proper ancestors and hence, by the above paragraph, $a_k$ is not
waiting. When $i > 0$, then all proper ancestors of $a_k$ 
are in safe positions (by hypothesis) 
and hence, by the inductive hypothesis, they are not waiting. 
Thus, by the above paragraph, $a_k$ is not waiting. 
 \end{proof}

\noindent
To show Thm.~\ref{loop-relation-2}, we need the following corollary of
Lemma~\ref{safe=notwaiting}. 

\begin{coro} \label{placed=instantiated}
Make the same assumptions as in Lemma~\ref{safe=notwaiting}.
If $Q_i=
a_1,\dots,a_n$ is $\pi_i$-robustly typed and
the atom $a_k$ selected in $Q_i\step Q_{i+1}$ has only ancestors
in safe positions, then
$\pi_i(k)=1$ (and hence $a_k$ is correctly typed in its input positions).
\end{coro}

\noindent
A permutation robustly typed query is called {\em well fed} if each
atom is robust or in a safe position. Note that if a
predicate $p$ can be shown to be robust using Lemma~\ref{terminates},
then all predicates $q$ with $p \sqsupset q$ are also robust. However,
this is a property of the method for showing robustness, not
of robustness itself. To simplify the proof of
Thm.~\ref{loop-relation-2}, we want to exclude the pathological
situation that $p$ is robust but some predicate $q$ with 
$p \sqsupset q$ is not.

\begin{defi}[well fed] \label{well-fed}
A $\pi$-robustly typed query is {\bf well fed} if for each of its atoms 
$p(\vect{s},\vect{t})$, either $p(\vect{s},\vect{t})$ is in a safe position for $\pi$, or
all predicates $q$ with $p \sqsupseteq q$ are robust.
A clause is well fed if its body is.
A program $P$ is {\bf well fed} if all of its clauses
are well fed and input-linear, and $P$ has input selectability.
\stopper \end{defi}

\noindent
The following proposition is a direct consequence of the
definition of the derived permutation (see  
Lemma~\ref{nicely-moded-persistence}).

\begin{propo}\label{well-fed-persistence}
Let $P$ and $Q$ be a well fed program and query, and $\xi$ a
derivation of $P \cup \{ Q \}$. Then each atom in each query 
in $\xi$ is either robust, or all its ancestors are in safe 
positions.
\end{propo}
 
\begin{example} \label{well-fed-ex}
The program in Fig.~\ref{ttl-fig} is 
well fed in both modes.
The program in Fig.~\ref{nqueens-fig} is
well fed in mode $M_1$.
It is not well fed in mode $M_2$, 
because it is
not \perm nicely moded in this mode: in the
second clause for {\tt sequence}, 
{\tt N1} occurs twice in an output position. 
\stopper \end{example}

\noindent
The following theorem reduces the problem of showing termination of
left-based derivations for a well fed program to showing termination of
LD-der\-i\-va\-tions for a corresponding robustly typed program. 

\begin{rmtheorem} \label{loop-relation-2}
Let $P$ and $Q$ be a well fed program and query, 
and $P'$ and $Q'$ a robustly typed program and query
corresponding 
to $P$ and $Q$.
If every LD-derivation of $P' \cup \{ Q' \}$
is finite, then
every left-based derivation of
$P \cup \{ Q \}$ is finite.
({\em Proof see Appendix})
\end{rmtheorem}

\noindent
Given that for the programs of Figs.~\ref{qs-fig},
\ref{permute-ok-fig}, 
\ref{ttl-fig} and \ref{nqueens-fig}, the
corresponding robustly typed programs terminate for robustly typed
queries, it follows by the above theorem that the original programs 
terminate for well fed queries.

For the program of Fig.~\ref{nqueens-fig}, 
our method can only show termination for the
mode $M_1$, but not for $M_2$, although the
program actually terminates for $M_2$ (provided the 
\block\ declarations are modified to allow for $M_2$). 

\section{Freedom from Errors Related to Built-ins} \label{error-sec}
 One problem with
built-ins is that their implementation may not be written in Prolog.
Thus, for the purposes of this paper, it is assumed that each built-in
is conceptually defined by
possibly infinitely many (fact) clauses~\cite{SS86}.
For example, there could be facts 
``\verb@0 is 0+0.@'', 
``\verb@1 is 0+1.@'', and so forth. 

To prove that a program is free from errors related to built-ins, we
require it to be permutation simply typed. This applies also to the
conceptual clauses for the built-ins.

Some built-ins produce an error if certain arguments have a wrong type,
and others produce an error if certain arguments are
insufficiently instantiated. For example, 
\hspace{0.2em} \verb@X is foo@ \hspace{0.2em} 
results in a type error and
\hspace{0.2em} \verb@X is V@ \hspace{0.2em} results in an instantiation error.

The approach described here aims at preventing 
instantiation and type errors for built-ins, for example arithmetic
built-ins, that require arguments to be ground.
It has been proposed~\cite{AL95} that these predicates be equipped with
delay declarations to ensure that they are only executed when the input is
{\em ground}. 
This has the advantage that one can reason about arbitrary
arithmetic expressions, 
 say \verb@qsort([1+1,3-8],M)@. The disadvantage is that 
\block\ declarations cannot be used. 
In contrast, we assume that the type of arithmetic built-ins
is the constant type $num$, rather than arithmetic
{\em expressions}. 
Then we show that \block\ declarations are
sufficient.
The following lemma is similar to and based on Lemma~27 \cite{AL95}.

\begin{rmlemma} \label{nonvar-enough}
Let 
$Q =  p_1(\vect{s}_1,\vect{t}_1),\dots,p_n(\vect{s}_n,\vect{t}_n)$ 
be a $\pi$-well typed query, where $p_i(\vect{S}_i,\vect{T}_i)$ 
is the type of $p_i$ for each 
$i \in \oneton$. 
Suppose, for some
$k \in \oneton$, $\vect{S}_k$ is a vector of constant types,
$\vect{s}_k$ is a vector of non-variable terms, and 
there is a substitution $\theta$ such that 
$\vect{t}_j \theta: \vect{T}_j$ for all $j$ with $\pi(j) < \pi(k)$. Then 
$\vect{s}_k : \vect{S}_k$. 
\end{rmlemma}

\begin{proof}
By Def.~\ref{well-typed}, 
$\vect{s}_k \theta : \vect{S}_k$, and thus $\vect{s}_k \theta$ is a vector of constants. Since
$\vect{s}_k$ is already a vector of non-variable terms, it follows that
$\vect{s}_k$ is a vector of constants and thus 
$\vect{s}_k \theta = \vect{s}_k$. Therefore $\vect{s}_k: \vect{S}_k$.
\end{proof}

\noindent
Note that if $\vect{s}_k$ is of type $\vect{S}_k$, then $\vect{s}_k$
is ground.
By Def.~\ref{simply-typed}, for every \perm simply typed query $Q$,
there is a $\theta$ such that $Q \theta$ is correctly typed in its
output positions. Thus by Lemma~\ref{nonvar-enough}, if the arithmetic
built-ins have type $num$ in all input positions, then it is enough to
have \block\ declarations such that these built-ins are only selected
when the input positions are non-variable.  This is stated in the
following theorem which is a consequence of
Lemma~\ref{nonvar-enough}.

\begin{rmtheorem} \label{block-enough}
Let $P$ be a \perm simply typed, input-linear program with input selectability and
$Q$ be a \perm simply typed query. Let $p$ be a predicate whose input
positions are all bound and of constant type. Then in any
delay-respecting derivation of $P \cup \{ Q \}$, an atom using $p$
will be selected only when its input arguments are correctly typed.
\end{rmtheorem}

\noindent
When we say that the 
input positions of a built-in are bound, we imply that the conceptual
clause heads have non-variable terms in those positions.

\begin{example}
For the program in Fig.~\ref{qs-fig} in mode $M_1$,
no delay-respecting derivation for a \perm simply typed query and this
program can result in an instantiation or type error 
related to the arithmetic built-ins.
\end{example}

\section{\block\ Declarations and Equality Tests} 
\label{improve-sec}
Runtime testing for instantiation has an overhead, and in
the case of built-ins, can only be realised by introducing
an auxiliary predicate (see Fig.~\ref{qs-fig}). Therefore in the
following two subsections, we describe ways of simplifying the
\block\ declarations of a program. An additional benefit is that 
in some cases, we can even ensure that arguments are ground, 
rather than just non-variable. We will see in
Subsec.~\ref{weaken-subsec} that this is useful in order to weaken the
restriction that every clause head must be input-linear.
We have postponed these considerations so far in order to
avoid making the main arguments of this paper unnecessarily
complicated.

\subsection{Avoiding \block\ Declarations for Permutation 
            Simply Typed Programs}\label{constant-drop-block-subsec}
In the program in Fig.~\ref{nqueens-fig}, there are no 
{\tt block} declarations and hence no auxiliary predicates for
\verb@<@, {\tt is} and \verb@=\=@. This is justified because the input
for those predicates is always provided by the
clause heads.
For example, it is not necessary to have a $\tt block$ declaration for 
\verb@<@ because when an atom using $\tt sequence$ is called, the
first argument of this atom is already ground. 
We show here how this intuition can be formalised. 
In the following definition, we consider a
set $\mathcal B$ containing the predicates
for which we want to omit the $\tt block$ declarations.

\begin{defi}[$\mathcal B$-ground] \label{sure-input-2}
Let $P$ be a permutation simply typed program and $\mathcal B$ a
set of predicates whose input positions are all of constant type.

A query is {\bf $\mathcal B$-ground}
if it is permutation simply typed 
and each atom using a predicate in $\mathcal B$ has ground terms in
its input positions.

An argument position $k$ of a predicate $p$ in $P$ is a 
{\bf $\mathcal B$-position} if there is a clause
$p(\vect{t}_0,\vect{s}_{n+1}) \leftarrow
p_1(\vect{s}_1,\vect{t}_1),\dots,p_n(\vect{s}_n,\vect{t}_n)$
in $P$ such that for some $i$ where $p_i \in \mathcal B$, 
some variable in $\vect{s}_i$ also occurs in position $k$ in 
$p(\vect{t}_0,\vect{s}_{n+1})$.

The program $P$ is {\bf $\mathcal B$-ground} if every 
$\mathcal B$-position of every predicate in $P$ is an input position
of constant type, and an atom $p(\vect{s},\vect{t})$, where 
$p\not\in \mathcal B$, is selectable 
only if it is non-variable in the $\mathcal B$-positions of $p$.
\stopper \end{defi}

\noindent
Note that since a constant type is, in particular, a {\em non-variable}
type, it is always possible to find \block\
declarations such that both the requirement on selectability in the above
definition and in Def.~\ref{selectable2} (\ref{max-sel}) are 
fulfilled.

\begin{example}\label{bground-ex}
The program in Fig.~\ref{nqueens-fig} is
$\mathcal B$-ground, where 
${\mathcal B} = \{ \verb@<@,\; {\tt is},\; \verb@=\=@\}$.
The first position of {\tt sequence}, the second position of 
{\tt safe\_aux}, and all positions of 
{\tt no\_diag} are $\mathcal B$-positions.
\stopper
\end{example}

\noindent
The following theorem says that for $\mathcal B$-ground programs,
the input of all atoms using predicates in $\mathcal B$ is always
ground.

\begin{rmtheorem} \label{dont-need-block-2}
  Let $P$ be a $\mathcal B$-ground, input-linear program with input
  selectability, 
  $Q$ a $\mathcal B$-ground query, and $\xi$ a 
  delay-respecting derivation of $P \cup \{Q\}$.  Then each query in
  $\xi$ is $\mathcal B$-ground.
\end{rmtheorem}

\begin{proof}
The proof is by induction on the length of $\xi$.
Let $Q_0 = Q$ and $\xi = Q_0;Q_1;\dots$.
The base case holds by the assumption that 
$Q_0$ is $\mathcal B$-ground.

Now consider some $Q_j$ where $j \geq 0$ and $Q_{j+1}$ exists. 
By Lemma~\ref{simply-typed-persistence} 
(\ref{simply-typed-persistence-main}),
$Q_{j}$ and $Q_{j+1}$ are permutation simply typed and hence
type-consistent in all argument positions.
 The induction hypothesis is that $Q_j$ is $\mathcal B$-ground. 

Let $p(\vect{u},\vect{v})$ be the selected atom, 
$C = p(\vect{t}_0,\vect{s}_{n+1}) \leftarrow
p_1(\vect{s}_1,\vect{t}_1),\dots,p_n(\vect{s}_n,\vect{t}_n)$ be the
clause and $\theta$ the MGU used in the step $Q_j \step Q_{j+1}$.
Consider an arbitrary $i \in \oneton$ such that $p_i \in \mathcal B$.

If $p \not\in \mathcal B$, then by the condition on selectability in
Def.~\ref{sure-input-2}, $p(\vect{u},\vect{v})$ is non-variable in the
$\mathcal B$-positions of $p$ and hence, since the 
$\mathcal B$-positions are of constant type, 
$p(\vect{u},\vect{v})$ is {\em ground} in the $\mathcal B$-positions
of $p$. If $p \in \mathcal B$, then $p(\vect{u},\vect{v})$ is ground in all input
positions by the induction hypothesis, and hence $p(\vect{u},\vect{v})$ is a
fortiori ground in all $\mathcal B$-positions of $p$. 

Thus it follows that $\vect{s}_i \theta$ is ground.  Since the
choice of $i$ was arbitrary and because of the induction hypothesis,
it follows that $Q_{j+1}$ is $\mathcal B$-ground.
\end{proof}

\noindent
In Thm.~\ref{dont-need-block-2}, the assumption that the predicates 
in $\mathcal B$ have input
selectability is redundant. Atoms using predicates in $\mathcal B$
are only selected when their input is ground, simply because their
input is ground at all times during the execution.

\begin{example}
  In the program in Fig.~\ref{nqueens-fig},
  there are no {\tt block} declarations,
  and hence no auxiliaries, for
  the occurrences of {\tt is}, \verb@<@ and \verb@=\=@,
but there are {\tt block} declarations on
$\tt safe\_aux$ and $\tt no\_diag$ that ensure the condition on
  selectability in Def.~\ref{sure-input-2}.
\stopper \end{example}

\subsection{Simplifying the \block\ Declarations Using Atoms in Safe Positions}  

By a simple observation, we can simplify the \block\ declarations for
predicates that are only used in atoms occurring in safe positions.
Consider a permutation robustly typed program $P$ with input selectability
and a permutation robustly typed query $Q$.  Suppose we have a predicate
$p$ such that for all $q$ with $q \sqsupseteq p$, all atoms using $q$
in $Q$ and clause bodies in $P$ are in safe positions.

Then by Lemma~\ref{safe=notwaiting}, in any left-based derivation
of $P \cup \{ Q \}$, an atom using $p$ is never waiting.  Thus,
the \block\ declarations do not delay the selection of atoms using
$p$. Suppose we modify $P$ by replacing the \block\ declaration for $p$
with the empty \block\ declaration. Then the modified program has the
same set of left-based derivations of $Q$ as the original program.
For example, the \block\ declaration for {\tt sequence} in the
program in Fig.~\ref{nqueens-fig} can be omitted.

\subsection{Weakening Input-Linearity of Clause Heads}\label{weaken-subsec}
The requirement that clause heads are input-linear
is needed to show the persistence of permutation nicely-modedness
(Lemma~\ref{nicely-moded-persistence}).
This is analogous
to the same statement restricted to
nicely-modedness~\cite[(Apt \& Luitjes, 1995, Lemma~11)]{AL95}. 
However, the clause head does not have to be input-linear 
when the statement is further restricted to {\em
LD}-resolvents~\cite[(Apt \& Pellegrini, 1994, Lemma~5.3)]{AP94}.  
The following example by
Apt (personal communication) demonstrates this difference.

\begin{example}\label{input-linear-ex}
Consider the program 
\begin{verbatim}
q(A).          r(1).          eq(A,A).
\end{verbatim}
where the mode is $\{ \tt q(\inp), r(\out), eq(\inp,\inp) \}$.
Note that ${\tt eq}/2$ is equivalent to the built-in $\mbox{\tt =}/2$.
This program is nicely moded but not input-linear. The query
\[
\tt q(X),\ r(Y),\ eq(X,Y)
\]
is nicely moded. The query $\tt q(X), r(X)$ 
is a resolvent of the above query, and it is {\em not} nicely moded.
\stopper \end{example}

\noindent
Requiring clause heads to be input-linear is undoubtedly a severe
restriction. It means that it is not possible to check two input
arguments for equality. However, this also indicates the reason
why in the above example, resolving \verb@eq(X,Y)@ is harmful: 
{\tt eq} is meant to be a check, clearly indicated by its mode $\tt
eq(\inp,\inp)$, but in the given derivation step, it actually is not a
check, since it binds variables.

It is easy to see that Lemma~\ref{nicely-moded-persistence} still
holds if Def.~\ref{nicely-moded} is weakened by allowing 
{\tt =} 
to be used in mode $\mbox{\tt =}(\inp,\inp)$, provided 
atoms using \verb@=@ are only resolved when both arguments are
ground. Resolving the \perm nicely moded query 
$Q_1, s \mbox{\tt =} t, Q_2$ selecting $s \mbox{\tt =} t$, where $s$ and $t$ are
ground, will yield the resolvent $Q_1, Q_2$, which is \perm
nicely moded.

The mode $\mbox{\tt =}(\inp,\inp)$ can be realised with a
delay declaration such that an atom $s \mbox{\tt =} t$ is selected only when $s$
and $t$ are ground. In \six, this can be done using the built-in 
{\tt when}~\cite{sicstus}. However we do not follow this line because
this paper focuses on \block\ declarations, and because it would
commit a particular occurrence of $s \mbox{\tt =} t$ to be a test in all modes in
which the program is used.

Nevertheless, there are at least two situations when clause heads that
are not input-linear can be allowed. First, one can exploit the fact
that atoms are in safe positions, and secondly, that the
arguments being checked for equality are of constant type.

In the first case, we assume left-based derivations.
We could allow for clause heads
$p(\vect{t},\vect{s})$ where a variable $x$ occurs in several input
positions, provided that
\begin{itemize}
\item
all occurrences of $x$ in $\vect{t}$
are in positions of ground type, and
\item
for each clause body and initial query for the program,
each atom using a predicate $q$ with $q \sqsupseteq p$ 
is in a safe position.
\end{itemize}
By Cor.~\ref{placed=instantiated}, it is then ensured that 
multiple occurrences of a variable in the input of a clause head 
implement an equality {\em check} between input arguments. 
Therefore, Lemmas~\ref{nicely-moded-persistence}, \ref{simply-typed-persistence}
and~\ref{robustly-typed-persistence} hold assuming this weaker
definition of ``input-linear''.

\begin{example}
\begin{sloppypar}
Consider the program in Fig.~\ref{permute-loops-fig} in mode
$\{ {\tt permute(\inp,\inp)}$, 
${\tt delete(\inp,\inp,\inp)}\}$.
This program is not input-linear. Nevertheless, the program can
be used in this mode provided that all arguments are of ground type
and calls to $\tt permute$ and 
$\tt delete$ are always in safe positions.
\stopper 
\end{sloppypar}
\end{example}

\noindent
In the second case, it is sufficient to assume delay-respecting
derivations. We can use Thm.~\ref{block-enough}.
This time, we have to allow for clause heads $p(\vect{t},\vect{s})$
where a variable $x$ occurs in several input positions,
provided that 
\begin{itemize}
\item
$x$ only occurs {\em directly} and in positions of constant type 
in $\vect{t}$, and
\item
an atom using $p$ is selectable only if these positions are non-variable.
\end{itemize}

\noindent
It is then ensured that when an atom $p(\vect{u},\vect{v})$ is
selected, $\vect{u}$ has constants in each position where $\vect{t}$ has $x$.

\begin{figure} 
\figrule
\begin{center}
\begin{minipage}[t]{6cm}
\small
\begin{verbatim}
:- block length(-,-).
length(L,N) :- 
  len_aux(L,0,N).

:- block less(?,-), less(-,?).
less(A,B) :- 
  A < B.
\end{verbatim}
\end{minipage}
\begin{minipage}[t]{6cm}
\small
\begin{verbatim}
:- block len_aux(?,-,?), 
         len_aux(-,?,-).
len_aux([],N,N).
len_aux([_|Xs],M,N) :-
  less(M,N),
  M2 is M + 1,
  len_aux(Xs,M2,N).
\end{verbatim}
\end{minipage}
\caption{The {\tt length} program \label{length-fig}}
\end{center}
\figrule
\end{figure}

\begin{example}
Consider the program shown in Fig.~\ref{length-fig}. It can be used in mode 
$\{ \tt length(\out,\inp),$
$\tt len\_aux(\out,\inp,\inp)\}$ (it is simply typed) 
in spite of the fact that 
$\tt len\_aux([],N,N)$ is not input-linear,  
using either of the two explanations above. 
The first explanation relies on all atoms using predicates 
$q \sqsupseteq {\tt len\_aux}$ being in safe positions. This is somewhat
unsatisfactory since imposing such a
restriction impedes modularity. Therefore, the second
explanation is preferable.
\stopper \end{example}

\section{Related Work} \label{related} \label{conclusion}
First of all, note that our work implicitly relies on 
previous work on termination for LD-derivations~\cite{A97,DD94}, since
we reduce the problem of termination of a program with \block\
declarations to the classical problem of termination for LD-derivations.

In using modes and types, we follow
{\bf Apt} and {\bf Luitjes}~\shortcite{AL95}, 
and also adopt their notation. 
They show occur-check freedom for nicely moded programs and 
non-floundering for well typed programs.
For arithmetic built-ins they require delay declarations
such that an atom is delayed until the arguments are ground. 
Such declarations are usually implemented not as
efficiently as \block\ declarations.  
For termination, they propose a
method limited to deterministic programs. 

{\bf Naish}~\shortcite{N92} gives good intuitive explanations (without proof) why
programs loop, which directed our own search for further ideas and
their formalisation.
Predicates are assumed to have a single mode. It is suggested that
alternative modes should be achieved by multiple versions of a
predicate.  This approach is quite common~\cite{AE93,AL95,EBC99} and
is also taken in Mercury~\cite{mercury}, where these versions are
generated by the compiler. 
While it is possible to take that approach, this is clearly a loss of
generality since two different versions of a predicate is not the
same thing as a single one which can be used in several modes.
Naish uses examples where, under the above assumption,
delay declarations are unnecessary. 
For {\tt permute},
if we only consider the mode $M_2$, then
the program in Fig.~\ref{permute-ok-fig} does not loop simply because 
no atom is ever delayed, and thus the program behaves as if there were no
delay declarations. In this case, the 
interpretation  that
one should ``place recursive calls last'' is misleading. 
If we only consider the mode $M_1$, then the version of 
Fig.~\ref{permute-ok-fig} is much less efficient than 
Fig.~\ref{permute-loops-fig}. In short, his discussion on
delay declarations lacks motivation when only one mode
is assumed. 

{\bf L\"uttringhaus-Kappel}~\shortcite{L93} proposes a method for generating control
automatically, and has applied it 
successfully to many
programs. However, rather than pursuing  a formalisation of some intuitive 
understanding of why programs loop, and imposing appropriate
restrictions on programs, he aims for a high degree of
generality. This has certain disadvantages. 

The
method only finds {\em acceptable} delay declarations, ensuring
that the most general selectable atoms have finite
SLD-trees. What is required however are {\em safe} delay declarations, 
ensuring that {\em instances} of most general selectable atoms have
finite SLD-trees. A {\em safe} program is a program for
which every acceptable delay declaration is safe. 
L\"uttringhaus-Kappel states that all programs he has considered are
safe, but he gives no hint as to how this might be shown in general. 

The delay declarations for
some programs such as {\tt quicksort} require an argument to be a
nil-terminated list before an atom can be selected. As
L\"uttringhaus-Kappel points out, ``in NU-Prolog [{\it or \six}]
it is not possible to express such conditions''. 
We have shown here that, with a knowledge of modes and types, \block\
declarations are sufficient. 

Furthermore, the method assumes arbitrary delay-respecting derivations and
hence does not work for programs where termination depends on
derivations being left-based.

{\bf Marchiori} and {\bf Teusink}~\shortcite{MT99} base termination on norms
and the {\em covering} relation between subqueries of a query. 
This is loosely related
to well-typedness. However, their results are not
comparable to ours because they assume a {\em local selection rule},
that is a rule that always selects an atom that was introduced in the
most recent step. 
No existing language
using a local selection rule 
(other than the LD selection rule)
is mentioned, and we are not aware
that there is one.
The authors state that
programs that do not use {\em speculative bindings} 
deserve further investigation, and that they expect 
any method for proving termination with {\em full} coroutining 
either to be very complex, or very restrictive in its applications.

{\bf Martin} and {\bf King}~\shortcite{MK97} ensure termination by imposing
a depth bound on the SLD tree. This is realised by a program
transformation introducing additional argument positions for each
predicate, which are counters for the depth of the computation. The
difficulty is of course to find an appropriate depth bound that does
not compromise completeness. It is hard to compare their work to ours
since they transform the programs substantially to obtain programs for
which it is easier to reason about termination, whereas 
we show termination for much more ``traditional'' programs.

Recently, {\bf Pedreschi} and {\bf Ruggieri}~\shortcite{PR99} have shown
that for programs that have no failing derivations, termination is
independent of the selection rule. They consider {\em guarded}
clauses, and the execution model is such that the evaluation of guards
is never considered as a failure. For example, even the 
$\tt quicksort$ program is non-failing in this sense, since the tests
\verb@leq(X,C)@ and \verb@grt(X,C)@ (see Fig.~\ref{qs-fig}) would be
guards. In contrast to the method presented in
Subsec.~\ref{nonspec-subsec}, they can 
 show termination for this program.

The verification methods used here can also be used to show that
programs are free from (full) unification, occur-check, and
floundering. These relatively straightforward generalisations of
previous results~\cite{AE93,AL95,AP94} are discussed in Smaus's 
PhD thesis~\shortcite{smaus-thesis}.

\section{Discussion and Future Work} \label{discussion} \label{efficiency}
We have presented verification methods for programs with
\block\ declarations. 
The verified properties were termination and 
freedom from 
errors related to built-ins.
These methods refine and formalise previous work in this area~\cite{AE93,AL95,N92}. 

In the introduction, we have said that this work has three distinctive
features: (a) assuming multiple modes, (b) using \block\ declarations, 
(c) formalising the ``default left-to-right'' selection rule. While
the significance of (a) can be argued (see below), at least features 
(b) and (c) mean that we are addressing existing programs and existing
language implementations. This is further strengthened by the fact
that, using the results of Sec.~\ref{improve-sec}, we can verify
programs where only some of the predicates are equipped with \block\
declarations. 

In the literature, we also find other types of delay declarations: 
In G\"odel~\cite{goedel}, delay
declarations can test for non-variableness of sub-arguments up to a
certain depth 
(e.g.~$\tt DELAY \ P([x|xs]) \ UNTIL \ NONVAR(xs)$) 
or for groundness of arguments; also, in theory, one can
consider delay declarations that test arguments for being instantiated 
to a list
or similar structure~\cite{L93}. Most of our results require that an 
atom is selected only if certain arguments are {\em at least}
non-variable, and so they trivially also hold for those delay
declarations. On the other hand, the results in 
Subsec.~\ref{well-fed-subsec} require that 
an atom  is definitely selectable whenever it is 
correctly typed in its input positions. We claim that this is a
natural requirement which should also be fulfilled by most programs
using other kinds of delay declarations, but to substantiate this
claim, we would have to specify precisely the delay declarations and
the underlying modes and types.

For proving termination, we have presented two approaches. The first
approach \cite{HKS-error} consists of two complementing methods based
on not {\em using} and not {\em making} speculative bindings,
respectively.  
For Figs.~\ref{ttl-fig} and~\ref{permute-ok-fig}, it turns
out that in one mode, the first method applies, and in the other
mode, the second method applies. This approach is simple to understand
and to apply. However it is rather limited. Termination cannot be
shown for the programs of Figs.~\ref{qs-fig} and~\ref{nqueens-fig}.

In the second approach~\cite{HKS-term}, we required programs to be
{\em permutation robustly typed}, a condition that ensures that no
call instantiates its own input.  In the next step, we identified when
a predicate is {\em robust}, which means that every delay-respecting
derivation for a query using the predicate terminates. Robust atoms
can be placed in clause bodies arbitrarily. Non-robust atoms must be
placed such that their input is sufficiently instantiated when they
are called.

Concerning built-ins, we have shown that even though some built-ins
require their input arguments to be ground, it is still sometimes
sufficient to use \block\ declarations. 

We have also considered how some of the \block\ declarations can be
omitted if it can be guaranteed that the instantiation tests they
implement are redundant. This is useful because even for
programs containing {\tt block} declarations, it is rare that 
{\em all} predicates have {\tt block} declarations. In particular,
it is awkward having to introduce auxiliary predicates to implement 
delay declarations for built-ins.

It is an ongoing discussion whether it is reasonable to assume
predicates that work in several modes~\cite{newsletter}. We have
argued that a formalism dealing with delay declarations should at
least {\em allow} for multiple modes.  This does not exclude in any
way other applications of delay declarations, such as
implementing the test-and-generate paradigm (coroutining).  As seen in
the program of Fig.~\ref{nqueens-fig}, 
our results apply to such programs as well.

The main purpose of this work is software development, and it is
envisaged that an implementation should take the form of a program
development tool.  The programmer would provide mode and type
information for the predicates in the program.  The tool would then
generate the \block\ declarations and try to reorder the atoms in
clause bodies so that the mode and type requirements are met.  Where
applicable, finding the free and bound positions, as well as the
level mapping used to prove robustness, should be done by the
tool.

\subsection*{Acknowledgements}
We are indebted to the anonymous referees whose comments helped
improve this paper. We also thank the referees of PLILP/ALP'98 and
LOPSTR'98, as well as Krzysztof Apt, Sandro Etalle, Lee Naish,
Salvatore Ruggieri and Sofie Verbaeten for useful discussions.

\appendix
\section{Proofs}

\noindent
{\bf Lemma~\ref{simply-typed-persistence}}
Let $Q =p_1(\vect{s}_1,\vect{t}_1),\dots,p_n(\vect{s}_n,\vect{t}_n)$ 
be a $\pi$-simply typed query and  
$C= p_k(\vect{v}_0,\vect{u}_{m+1}) 
\leftarrow 
q_1(\vect{u}_1,\vect{v}_1),\dots,q_m(\vect{u}_m,\vect{v}_m)$ 
a  $\rho$-simply typed, input-linear clause where 
$\vars(C) \cap \vars(Q) = \emptyset$. 
Suppose that for some $k \in \oneton$, 
$\vect{s}_k$ is non-variable in all 
bound input positions,
and $\theta$ is the MGU of $p_k(\vect{s}_k,\vect{t}_k)$ and 
$p_k(\vect{v}_0,\vect{u}_{m+1})$. Then 
\begin{enumerate}
\item
there exist substitutions $\theta_1$, $\theta_2$ such that
$\theta = \theta_1\theta_2$ and
  \begin{enumerate}
  \item 
  $\vect{v}_0\theta_1 = \vect{s}_k$ and
  $\dom(\theta_1) \subseteq \vars(\vect{v}_0)$,
  \item 
  $\vect{t}_k\theta_2 = \vect{u}_{m+1} \theta_1$ and 
  $\dom(\theta_2) \subseteq \vars(\vect{t}_k)$,
  \end{enumerate}
\item 
$\dom(\theta) 
\subseteq \vars(\vect{t}_k) \cup \vars(\vect{v}_0)$,
\item  
$\dom(\theta) \cap \vars(\vect{t}_1,\dots,\vect{t}_{k-1},
\vect{v}_1,\dots,\vect{v}_m,
\vect{t}_{k+1},\dots,\vect{t}_n) =
\emptyset$,
\item 
the resolvent of $Q$ and $C$ with selected atom 
$p_k(\vect{s}_k,\vect{t}_k)$ is $\varrho$-simply typed, where $\varrho$
is the derived permutation
(see Lemma~\ref{nicely-moded-persistence}).
\end{enumerate}

\begin{proof}
By assumption $\vect{s}_k$ is non-variable in all bound positions, and 
$\vect{v}_0$ is a linear vector having flat terms in all
bound positions, and   
variables in all other positions.
Thus there is a substitution $\theta_1$ such that 
$\vect{v}_0 \theta_1 = \vect{s}_k$ and 
$\dom(\theta_1) \subseteq \vars(\vect{v}_0)$, which shows (\ref{dom1}).

Since $\vect{t}_k$ is a linear vector of variables, there is
a substitution $\theta_2$ such that 
$\dom(\theta_2) \subseteq \vars(\vect{t}_k)$ and
$\vect{t}_k \theta_2 = \vect{u}_{m+1} \theta_1$,
which shows (\ref{dom2}).

Since $Q$ is $\pi$-nicely moded, 
$\vars(\vect{t}_k) \cap \vars(\vect{s}_k) = \emptyset$, and therefore
$\vars(\vect{t}_k) \cap \vars(\vect{v}_0 \theta_1) = \emptyset$.
Thus it follows by (\ref{dom2}) that $\theta = \theta_1 \theta_2$ is a unifier of
$p_k(\vect{s}_k,\vect{t}_k)$ and $p_k(\vect{v}_0,\vect{u}_{m+1})$.
(\ref{dom_total}) follows from (\ref{dom1}) and (\ref{dom2}), and
(\ref{orig_var}) follows from (\ref{dom_total}) because of linearity.

By Lemma~\ref{nicely-moded-persistence} and~\ref{well-typed-persistence},
the resolvent is $\varrho$-nicely moded and $\varrho$-well typed. 
By (\ref{orig_var}), the vector of the 
output arguments of the resolvent is a linear vector of variables,
and hence (\ref{simply-typed-persistence-main}) follows.
 \end{proof}

\noindent
{\bf Lemma~\ref{robustly-typed-persistence}}
Let $Q =  
p_1(\vect{s}_1,\vect{t}_1),\dots,p_n(\vect{s}_n,\vect{t}_n)$
a $\pi$-robustly typed query and 
$C = 
p_k(\vect{v}_0,\vect{u}_{m+1}) \leftarrow
q_1(\vect{u}_1,\vect{v}_1),\dots,q_m(\vect{u}_m,\vect{v}_m)$
a $\rho$-robustly typed, input-linear clause where
$\vars(Q) \cap \vars(C) = \emptyset$. 
Suppose that for some $k \in \oneton$, 
$p_k(\vect{s}_k,\vect{t}_k)$ is non-variable in all bound input 
positions
and $\theta$ is the MGU of $p_k(\vect{s}_k,\vect{t}_k)$ and 
$p_k(\vect{v}_0,\vect{u}_{m+1})$. 

Then the resolvent of $Q$ and $C$ with selected atom $p_k(\vect{s}_k,\vect{t}_k)$ is 
$\varrho$-robustly typed, where $\varrho$ is the derived permutation
(see  Lemma~\ref{nicely-moded-persistence}). Moreover
$\dom(\theta) \cap \vars(\vect{s}_k) = \emptyset$.

\begin{proof} 
We show how $\theta$ is computed, where we consider three stages.  In
the first, $\vect{s}_k$ and $\vect{v}_0$ are unified. In the second,
the output positions are unified where the bindings go from $C$ to
$Q$.  In the third, the output positions are unified where the
bindings go from $Q$ to $C$.  Figure~\ref{proof-illus} illustrates
which variables are bound in each stage. 
The first three parts of the proof correspond to
the three stages of the unification.

\vspace{1ex}
\noindent
{\bf Part 1} (unifying $\vect{s}_k$ and $\vect{v}_0$). 
By Def.~\ref{robustly-typed}, 
$\vect{v}_0$ is a vector of flat terms, where
$\vect{v}^{\sf f}_0$ is a vector of variables, and by
assumption, $\vect{v}_0$ is linear. 
By assumption,
$\vect{s}^{\sf b}_k$ is a vector of non-variable terms and, 
since $\vars(C) \cap \vars(Q) = \emptyset$,
$\vars(\vect{v}_0) \cap
\vars(\vect{s}_k) = \emptyset$.
Thus there is a 
(minimal) substitution $\theta_1$ such that 
$\vect{v}_0 \theta_1 = \vect{s}_k$.
We show that the following hold:
\begin{itemize}
\setlength{\itemindent}{1.3em}
\item[(1a)]
$\dom(\theta_1) \cap \vars(\vect{s}_k) = \emptyset$. 
\item[(1b)]
$\dom(\theta_1) \cap 
\vars(\vect{v}_1,\dots,\vect{v}_m,
\vect{t}_1,\dots,\vect{t}_n) = \emptyset$.
\item[(1c)]
Let $x$ be a variable occurring directly in a position of type $\tau$ in 
$\vect{u}^{\sf b}_{m+1} \theta_1$. Then 
$x \notin \vars(\vect{s}_k)$. Moreover, $x$
can only occur in
$\vect{v}_1,\dots,\vect{v}_m, 
\vect{t}_1,\dots,\vect{t}_n$ 
in a bound position of type $\tau$,  and the occurrence must be direct. 
\item[(1d)]
$\vars(\vect{u}_{m+1} \theta_1) \cap \vars(\vect{t}_k)
= \emptyset$.
\end{itemize}
(1a) holds by the construction of $\theta_1$.

\setlength{\unitlength}{0.71cm}
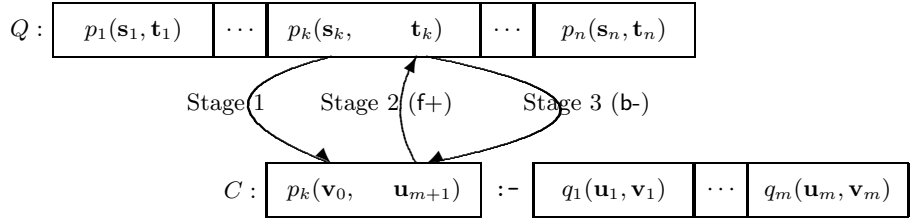
\begin{figure}
\begin{picture}(17,4)
\put(4,0){\makebox(1,1){$C:$}}
\put(5,0){\framebox(4,1){}}
\put(5,0){\makebox(2,1){$p_k(\vect{v}_0,$}}
\put(7,0){\makebox(2,1){$\vect{u}_{m+1})$}}
\put(9,0){\makebox(1,1){\tt :-}}
\put(10,0){\framebox(3,1){$q_1(\vect{u}_1,\vect{v}_1)$}}
\put(13,0){\framebox(1,1){$\cdots$}}
\put(14,0){\framebox(3,1){$q_m(\vect{u}_m,\vect{v}_m)$}}
\put(0,3){\makebox(1,1){$Q:$}}
\put(1,3){\framebox(3,1){$p_1(\vect{s}_1,\vect{t}_1)$}}
\put(4,3){\framebox(1,1){$\dots$}}
\put(5,3){\framebox(4,1){}}
\put(5,3){\makebox(2,1){$p_k(\vect{s}_k,$}}
\put(7,3){\makebox(2,1){$\vect{t}_k)$}}
\put(9,3){\framebox(1,1){$\dots$}}
\put(10,3){\framebox(3,1){$p_n(\vect{s}_n,\vect{t}_n)$}}
\qbezier(6.2,3)(3.1,2.2)(6.2,1)
\thicklines \put(6.2,1){\vector(1,-1){0}} \thinlines
\put(3.5,2){\parbox{2\unitlength}{Stage 1}}
\qbezier(7.8,3)(7.2,2)(7.8,1)
\thicklines \put(7.8,3){\vector(1,2){0}} \thinlines
\put(6,2){\parbox{3.5\unitlength}{Stage 2 ({\sf f+})}}
\qbezier(8,3)(13,2)(8,1)
\thicklines \put(8,1){\vector(-2,-1){0}} \thinlines
\put(9.8,2){\parbox{3\unitlength}{Stage 3 ({\sf b-})}}
\end{picture}
\caption{Data flow in the unification \label{proof-illus}}
\end{figure}

(1b) holds since by Def.~\ref{robustly-typed} and since $C$ is input-linear,
$\vect{v}_0,\dots,\vect{v}_m,
\vect{t}_1,\dots,\vect{t}_n$ is linear.
  
Let $x$ be a variable occurring directly in a position of type $\tau$
in $\vect{u}^{\sf b}_{m+1} \theta_1$. Let $y$ be the variable in the
same position in $\vect{u}^{\sf b}_{m+1}$.
Suppose, for the purpose of deriving a contradiction, that 
$y \in \vars(\vect{v}_0)$. Then by
Def.~\ref{robustly-typed},  
$y$ occurs {\em directly} in $\vect{v}^{\sf b}_0$, 
and since $\vect{s}^{\sf b}_k$ is a vector of non-variable terms,
$y \theta_1$ is not a variable, which is a contradiction.
Therefore $y \not\in \vars(\vect{v}_0)$. Hence 
$y \not\in \dom(\theta_1)$ and thus $x = y$ and 
$x \notin \vars(\vect{s}_k)$. Furthermore it follows by 
Def.~\ref{robustly-typed} that $x$ can only occur in 
$\vect{v}_1,\dots,\vect{v}_m,
\vect{t}_1,\dots,\vect{t}_n$ in a bound position of type
$\tau$, and the occurrence must be direct.
Thus (1c) holds.

Since $Q$ is \perm nicely moded,
$\vars(\vect{s}_k) \cap \vars(\vect{t}_k) = \emptyset$ and hence
$\ran(\theta_1) \cap \vars(\vect{t}_k) = \emptyset$. 
Thus (1d) holds.

\vspace{1ex}
\noindent
{\bf Part 2}
(unifying $\vect{t}_k$ and $\vect{u}_{m+1} \theta_1$ in each
position where either the argument in $\vect{t}_k$ is a variable, 
or the arguments in
$\vect{t}_k$ and $\vect{u}_{m+1} \theta_1$ are both
non-variable). 
Note that this includes all positions in $\vect{t}^{\sf f}_k$
and $\vect{u}^{\sf f}_{m+1} \theta_1$, but may also include positions
in $\vect{t}^{\sf b}_k$ and $\vect{u}^{\sf b}_{m+1} \theta_1$.
Since, by (1b), $\vect{t}_k \theta_1 = \vect{t}_k$, 
Part 2 covers precisely the output positions where the binding ``goes from 
 $\vect{u}_{m+1} \theta_1$ to $\vect{t}_k \theta_1$'' (see
 Fig.~\ref{proof-illus}).
We denote by $\vect{t}^{\sf f+}_k$ the projection of
$\vect{t}_k$ onto the positions where the argument in $\vect{t}_k$ 
is a variable, or the arguments in
$\vect{t}_k$ and $\vect{u}_{m+1} \theta_1$ are both
non-variable, and by $\vect{t}^{\sf b-}_k$ the projection
onto the other positions, and likewise for 
$\vect{u}_{m+1} \theta_1$.

By (1d), 
$\vars(\vect{u}_{m+1}^{\sf f+} \theta_1) \cap 
 \vars(\vect{t}_k^{\sf f+}) = \emptyset$.
Thus there is a minimal
substitution $\theta'$ such that 
$\vect{t}^{\sf f+}_k\theta' = 
\vect{u}^{\sf f+}_{m+1} \theta_1$.
Let $\theta_2 = \theta_1 \theta'$. Then
by (1b), $\vect{t}^{\sf f+}_k\theta_2 = \vect{u}^{\sf f+}_{m+1} \theta_2$.
We show the following:

\begin{itemize}
\setlength{\itemindent}{1.3em}
\item[(2a)]
$\dom(\theta_2) \cap \vars(\vect{s}_k) = \emptyset$. 
\item[(2b)]
$\dom(\theta_2) \cap 
\vars(\vect{v}_1,\dots,\vect{v}_m, 
\vect{t}_1,\dots,\vect{t}_{k-1},
\vect{t}^{\sf b-}_k,\vect{t}_{k+1},\dots,\vect{t}_n) = \emptyset$. 
\item[(2c)]
Let $x$ be a variable occurring directly in a position of type $\tau$ in 
$\vect{u}^{\sf b-}_{m+1} \theta_2$. Then 
$x \notin \vars(\vect{s}_k)$. Moreover, $x$ can only occur in 
$\vect{v}_1,\dots,\vect{v}_m,$ 
$\vect{t}_1,\dots,
\vect{t}_{k-1},$
$\vect{t}^{\sf b-}_k,$
$\vect{t}_{k+1},\dots,\vect{t}_n$ 
in a bound position of type $\tau$,  and the occurrence must be direct.
\item[(2d)]
$\vars(\vect{u}_{m+1} \theta_2) \cap \vars(\vect{t}^{\sf b-}_k)
= \emptyset$.
\end{itemize}

\noindent
Since $\vars(\vect{s}_k) \cap  \vars(\vect{t}_k) = \emptyset$,
$\dom(\theta') \cap \vars(\vect{s}_k) = \emptyset$.
This and (1a) imply (2a).

 (2b) holds because (1b) holds and 
$\vect{v}_1,\dots,\vect{v}_m, 
\vect{t}_1,\dots,\vect{t}_n$ is linear.

By (1d), 
$\dom(\theta') \cap \vars(\vect{u}_{m+1}^{\sf b-} \theta_1) = \emptyset$.
This together with (1c) implies (2c).
Furthermore, because of the linearity of $\vect{t}_k$, (2d) follows.

\vspace{1ex}
\noindent
{\bf Part 3} (unifying $\vect{t}^{\sf b-}_k$ and 
$\vect{u}^{\sf b-}_{m+1} \theta_2$). 
By (1d), $\dom(\theta') \cap \vars(\vect{u}_{m+1}^{\sf b-} \theta_1) =
\emptyset$, and thus 
$\vect{u}^{\sf b-}_{m+1} \theta_2 = 
\vect{u}^{\sf b-}_{m+1} \theta_1$. Therefore,
by the definition of the superscript $\sf b-$ in Part~2,
$\vect{u}^{\sf b-}_{m+1} \theta_2$ 
is a vector of variables. 
By (2d), 
$\vars(\vect{u}^{\sf b-}_{m+1} \theta_2) \cap
\vars(\vect{t}^{\sf b-}_{k}) = \emptyset$, 
so that 
there is a minimal substitution $\theta''$ such that 
$\vect{u}^{\sf b-}_{m+1} \theta_2 \theta'' = 
 \vect{t}^{\sf b-}_{k}$.
Let $\theta_3 = \theta_2 \theta''$.
Then, by (2b), we have $\vect{u}^{\sf b-}_{m+1} \theta_3 = 
 \vect{t}^{\sf b-}_{k}\theta_3$.
We show (3a) and (3b).
\begin{itemize}
\setlength{\itemindent}{1.3em}
\item[(3a)]
$\dom(\theta_3) \cap \vars(\vect{s}_k) = \emptyset$. 
\item[(3b)]
$(\vect{v}_1,\dots,\vect{v}_m, 
\vect{t}_1,\dots,\vect{t}_{k-1},\vect{t}_{k+1},\dots,\vect{t}_n) \theta_3$ 
is linear and has flat 
type-consistent terms in all bound positions and variables in all free
positions. 
\end{itemize}

\noindent
By (2c), 
$\dom(\theta'') \cap \vars(\vect{s}_k) = \emptyset$. This and (2a)
imply (3a).

Suppose $x$ is a variable in $\vect{u}^{\sf b-}_{m+1} \theta_2$
occurring in a position $i$ of type $\tau$, and
$x$ also occurs in 
$\vect{v}_1,\dots,\vect{v}_m,
\vect{t}_1,\dots,\vect{t}_{k-1},\vect{t}_{k+1},\dots,\vect{t}_n$. 
By  (2c), the latter occurrence of $x$ is in a bound position of type
$\tau$, and is the {\em only} occurrence of $x$ in 
$\vect{v}_1,\dots,\vect{v}_m,
\vect{t}_1,\dots,\vect{t}_{k-1},\vect{t}_{k+1},\dots,\vect{t}_n$.
Let $I$ be the set of positions where $x$ occurs in 
$\vect{u}^{\sf b-}_{m+1} \theta_2$, and
let $T$ be the set of terms occurring in  
$\vect{t}^{\sf b-}_k$ in positions in $I$.
Then $T$ is a
set of variable-disjoint, flat terms. 
Therefore their most general
common instance $x \theta''$ is a flat term and $x \theta''$ is
type-consistent with respect to $\tau$.
Moreover, since $(\vect{v}_1,\dots,\vect{v}_m,
\vect{t}_1,\dots,\vect{t}_{k-1},\vect{t}^{\sf b-}_k,
\vect{t}_{k+1},\dots,\vect{t}_n)$ is linear, we have
$\vars(x \theta'') \cap \vars(\vect{v}_1,\dots,\vect{v}_m,
\vect{t}_1,\dots,\vect{t}_{k-1},\vect{t}_{k+1},\dots,\vect{t}_n) = \emptyset$
and therefore it follows that 
$(\vect{v}_1,\dots,\vect{v}_m,
\vect{t}_1,\dots,\vect{t}_{k-1},\vect{t}_{k+1},\dots,\vect{t}_n)
\theta''$ is a linear vector of type-consistent terms.
This and (2b) imply (3b).

\vspace{1ex}
\noindent
{\bf Part 4:}
Defining $\theta = \theta_3$ it follows that
$p_k(\vect{s}_k,\vect{t}_k) \theta =  p_k(\vect{v}_0,\vect{u}_{m+1}) \theta$.
By (3b) and Lemmas~\ref{nicely-moded-persistence} and~\ref{well-typed-persistence}, 
the resolvent of $Q$ and $C$ is $\varrho$-robustly typed. 
By (3a), we have $\vect{s}_k \theta = \vect{s}_k$.
 \end{proof}

\noindent
{\bf Theorem~\ref{loop-relation}}
Let $P$ be a non-speculative program with input selectability 
and $P'$ a simply typed program
corresponding to $P$.
Let $ Q$ be a
\perm simply typed query and $ Q'$ a simply typed 
query corresponding to $Q$. 
If there is an infinite delay-respecting derivation of
$P \cup \{ Q \}$, 
then there is  an infinite LD-derivation of
$P' \cup \{ Q' \}$.

\begin{proof}
For simplicity assume that $Q$ and each clause body
do not contain two identical atoms.
Let $Q_0 = Q$, 
$\theta_0 = \emptyset$ and 
$\xi = 
\langle Q_0, \theta_0 \rangle; 
\langle Q_1, \theta_1 \rangle; 
\dots$ 
be a
delay-respecting derivation of $P \cup \{ Q \}$. 
The idea is to construct an LD-derivation $\xi'$ of
$P' \cup \{ Q' \}$ such that whenever $\xi$ uses a clause $C$, then
$\xi'$ uses the corresponding clause $C'$ in $P'$.
It will then turn out that if $\xi'$ is finite, 
$\xi$ must also be finite.

We call an atom $a$ {\bf resolved in $\xi$ at $i$} if
$a$ occurs in $Q_i$ but not in $Q_{i+1}$. We call  
$a$ {\bf resolved in $\xi$} if for some $i$, $a$ is resolved in $\xi$ at $i$. 
Let $Q'_0 = Q'$ and $\theta'_0 = \emptyset$. 
We construct an LD-derivation
$\xi' = 
\langle Q'_0, \theta'_0 \rangle; 
\langle Q'_1, \theta'_1 \rangle; 
\dots$ 
of $P' \cup \{ Q' \}$
showing that for each $i\geq 0$ 
the following hold:
\begin{enumerate}
\item[(1)]
If $q(\vect{u},\vect{v})$ is an atom in $Q'_i$ that is
 not resolved in $\xi$, then 
$\vars(\vect{v} \theta'_i) \cap \dom(\theta_j) = \emptyset$ for all $j\geq 0$.
\item[(2)]
Let $x$ be a variable such that, for
some $j\geq 0$, $x \theta_j = f(\dots)$. Then 
$x \theta'_i$ is either a variable or $x \theta'_i = f(\dots)$.
\end{enumerate}  

\noindent
We first show these properties for $i = 0$. Let 
$q(\vect{u},\vect{v})$ be an atom in $Q'_0$ that is
 not resolved in $\xi$.
Since $\theta_0' = \emptyset$, $\vect{v} \theta'_0 = \vect{v}$. 
Furthermore, by
Cor.~\ref{simply-typed-persistence-ld} and Cor.~\ref{star-dagger} 
and since
$q(\vect{u},\vect{v})$ is not resolved in~$\xi$, we have
$\vect{v} \theta_{j} = \vect{v}$ for all~$j$. Thus (1) holds.
(2) holds because $\theta'_0 = \emptyset$.

Now assume that for some $i$, 
$\langle Q'_i, \theta'_i \rangle$
is defined, $Q'_i$ is not empty, and (1) and (2) hold.
 Let $p(\vect{s},\vect{t})$ be the leftmost atom of $Q'_i$.
We define a derivation step 
$\langle Q'_i, \theta'_i \rangle;\langle Q'_{i+1}, \theta'_{i+1} \rangle$
with  $p(\vect{s},\vect{t})$ as the selected atom,
and show that (1) and (2) hold for $\langle Q'_{i+1}, \theta'_{i+1} \rangle$.

{\bf Case 1:}
$p(\vect{s},\vect{t})$ is resolved in $\xi$ at $l$ for some $l$.
Consider the simply typed clause 
$C' = h \leftarrow B'$ corresponding 
to the {\em uniquely renamed} clause (using the same renaming)
used in $\xi$ to resolve 
$p(\vect{s},\vect{t})$.
Since
$p(\vect{s},\vect{t})$ is resolved in $\xi$ at $l$, 
$p(\vect{s},\vect{t}) \theta_l$ is non-variable in all bound
input positions. Thus each bound input position of
$p(\vect{s},\vect{t})$ must be filled by a non-variable term or a
variable $x$ such that $x \theta_l = f(\dots)$ for some $f$. 
 Moreover, $p(\vect{s},\vect{t}) \theta'_i$ must
have non-variable terms in all bound input positions since 
$Q'_i \theta'_i$ is well typed.
Thus it follows by (2) that in each bound input
position, 
$p(\vect{s},\vect{t}) \theta'_i$ has the same top-level functor as 
$p(\vect{s},\vect{t}) \theta_l$, 
and since $h$ has flat
terms in the bound input positions, there is an MGU 
$\phi'_i$ of $p(\vect{s},\vect{t}) \theta'_i$ and $h$.
We use $C'$ for the step
$\langle Q'_i, \theta'_i \rangle
\step
\langle Q'_{i+1}, \theta'_{i+1} \rangle$.

We must show that (1) and (2) hold for $i+1$. 
Consider an atom 
$q(\vect{u},\vect{v})$ in $Q'_i$ other than $p(\vect{s},\vect{t})$.
By
Lemma~\ref{simply-typed-persistence} (\ref{orig_var}),
$\vars(\vect{v} \theta'_i) \cap \dom(\phi'_i) = \emptyset$.
Thus for the atoms in $Q'_{i+1}$
that occur already in $Q'_i$, (1) is maintained.
Now consider an atom 
$q(\vect{u},\vect{v})$ in $B'$ that is not 
resolved in~$\xi$. By Cor.~\ref{star-dagger},
$\vect{v} \theta'_{i+1} = \vect{v}$. Since 
$q(\vect{u},\vect{v})$ is not resolved in $\xi$, 
for all $j > l$ we have that 
$q(\vect{u},\vect{v})$ occurs in $Q_j$ and thus 
by Cor.~\ref{star-dagger},
$\vect{v} \theta_{j} = \vect{v}$. Thus (1) follows.
(2) holds since it holds for $i$ and $p(\vect{s},\vect{t})$ 
is resolved using the same clause
head as in $\xi$.

{\bf Case 2:}
$p(\vect{s},\vect{t})$ is not resolved in $\xi$.
Since $P'$ is non-speculative, there is a (uniquely renamed) clause 
$C' = h \leftarrow B'$ in $P'$ such that $h$ and 
$p(\vect{s},\vect{t}) \theta'_i$ have an MGU
$\phi'_i$. We use $C'$ for the step
$\langle Q'_i, \theta'_i \rangle
\step
\langle Q'_{i+1}, \theta'_{i+1} \rangle$.

We must show that (1) and (2) hold for $i+1$.
Consider an atom 
$q(\vect{u},\vect{v})$ in $Q'_i$ other than $p(\vect{s},\vect{t})$.
By
Lemma~\ref{simply-typed-persistence} (\ref{orig_var}), 
$\vars(\vect{v} \theta'_i) \cap \dom(\phi'_i) = \emptyset$.
Thus for the atoms in $Q'_{i+1}$
that occur already in $Q'_i$, (1) is maintained.
Now consider an atom 
$q(\vect{u},\vect{v})$ in $B'$. Clearly $q(\vect{u},\vect{v})$ 
is not resolved in $\xi$. Since 
$\vars(C') \cap \vars(Q_j \theta_j) = \emptyset$ for all 
$j$ and since by Cor.~\ref{star-dagger}, we have
$\vect{v} \theta'_{i+1} = \vect{v}$, (1) holds for $i+1$.

By (1) for $i$, we have
$\vars(\vect{t} \theta'_i) \cap \dom(\theta_j) = \emptyset$ for all $j$.
By Lemma~\ref{simply-typed-persistence} (\ref{dom_total}), we have
$\dom(\phi'_i) 
\subseteq \vars(\vect{t} \theta'_i) \cup \vars(C')$. 
Thus we have $\dom(\phi'_i) \cap \dom(\theta_j) = \emptyset$ for all $j$.
Moreover, (2) holds for $i$.
Thus (2) holds for $i+1$.

Since this construction can only terminate when the query is empty, 
either $Q_n'$ is empty for some $n$, or $\xi'$ is infinite.

Thus 
 we show that if $\xi'$ is finite, then every atom resolved
in $\xi$ is also resolved in $\xi'$. So let $\xi'$ be
finite of length $n$.
Assume for the sake of
deriving a contradiction that $j$ is the smallest number such that the
atom $a$ selected in 
$\langle Q_j, \theta_j \rangle \step
\langle Q_{j+1}, \theta_{j+1} \rangle$ is never selected in
$\xi'$. Then
$j\not= 0$ since $Q_0$ and $Q'_0$ are permutations of each other and all
atoms in $Q'_0$ are eventually selected in $\xi'$. Thus there must be a
$k< j$ such that $a$ does not occur in $Q_{k}$ but does occur in
$Q_{k+1}$.  Consider the atom $b$ selected in
$\langle Q_k, \theta_k \rangle \step
\langle Q_{k+1}, \theta_{k+1} \rangle$.  Then
by the assumption that $j$ was minimal, 
$b$ must be the selected atom in 
$\langle Q'_i, \theta'_i \rangle \step
\langle Q'_{i+1}, \theta'_{i+1} \rangle$
for some $i \leq n$.  Hence $a$ must occur in $Q'_{i+1}$,
since the clause used to resolve $b$ in $\xi'$ is a simply typed clause
corresponding to the clause used to resolve $b$ in $\xi$. 
Thus $a$ must occur in
$Q_n'$, contradicting that $\xi'$ 
terminates with the empty query.

Thus $\xi$ can only be infinite if $\xi'$ is also infinite.
\end{proof}

\noindent
{\bf Lemma~\ref{needs-feeding}}
Let $P$ be a robustly typed, input-linear program with 
input selectability and
 $F,b,H$ a robustly typed query where 
$b$ is a robust atom.  A delay-respecting derivation of 
$P \cup \{F,b,H\}$ can
have infinitely many $b$-steps only if it has infinitely many
$a$-steps, for some $a \in F$. 
\begin{proof}
In this proof, 
by an {\bf $F$-step} we mean  an  $a$-step, for some $a \in F$;
 likewise we define an {\bf $H$-step}.
By Cor.~\ref{consumer-affects-not}, no $H$-step
can instantiate any descendant of $F$ or $b$. Thus the $H$-steps can
be disregarded, and without loss of generality, we assume $H$ is empty. 
Suppose $\xi$ is a delay-respecting derivation for $P\cup\{F,b\}$ containing 
only finitely many $F$-steps. 

All $F$-steps are contained in a finite
prefix of $\xi$. Moreover, by Cor.~\ref{consumer-affects-not}, no  $b$-step can
instantiate any descendant of $F$. Therefore, we can repeatedly apply
the Switching Lemma~\cite[(Lloyd, 1987, Lemma 9.1)]{L87} to this prefix of $\xi$ to obtain a
delay-respecting derivation
\[
\xi_2 = 
\langle F,\!b,\; \emptyset \rangle 
\steps
\langle F',\!b,\; \rho \rangle
\step
\xi'
\]
such that 
$\langle F,\!b,\; \emptyset \rangle 
\steps
\langle F',\!b,\; \rho \rangle$
contains only $F$-steps and
$\langle F',\!b,\; \rho \rangle
\step 
\xi'$ 
contains only $b$-steps.
Now construct the delay-respecting derivation
\[
\xi_3 = 
\langle b, \rho \rangle
\step
\xi'_3
\]
by removing the prefix $F'$ in each query in 
$\langle F',\!b,\; \rho \rangle
\step
\xi'
$.

By Lemma~\ref{robustly-typed-persistence}, $(F',b) \rho$ is robustly
typed. Thus by Lemma~\ref{sigma-exists}, there exists a substitution
$\sigma$ such that $b \rho \sigma$ is robustly typed, and
$\dom(\sigma) = V$, where $V$ is the set of variables occurring 
in the output arguments of $F' \rho$. 

By Cor.~\ref{consumer-affects-not}, no $b$-step in $\xi_2$, and
hence no derivation step in $\xi_3$, can instantiate a variable in
$V$. Since $\dom(\sigma) = V$, it thus follows that we
can construct a delay-respecting derivation
\[
\xi_4 = 
\langle b, \rho \sigma \rangle
\step
\xi'_3 \sigma
\]
by applying $\sigma$ to each query in $\xi_3$.

Since $b \rho \sigma$ is a robustly typed query and $b$ is robust, $\xi_4$ is finite.
Therefore $\xi_3$, $\xi_2$,  and finally
$\xi$ are finite.
\end{proof}

\setcounter{equation}{0}

\noindent
{\bf Lemma~\ref{terminates}}
Let $P$ be a robustly typed, input-linear program with input 
selectability and~$p$ a predicate in $P$. 
Suppose all predicates $q$ with $p \sqsupset q$ are 
robust, and all clauses defining predicates 
$q \in [p]_\approx$ are well-recurrent with respect to some level
mapping $|.|$. 
Then $p$ is robust. 

\begin{proof}
If $a$ is an atom using a predicate in $[p]_\approx$ such that the set
$S = \{|a \theta| \mid a \theta \in \iatoms\}$ 
is non-empty and bounded, we define
$||a|| = sup(S)$.
Thus, for each atom $a$ and substitution $\theta$
such that  $||a||$ and $||a\theta||$ are defined 
\begin{equation}
||a \theta|| \leq ||a||
\label{only-smaller}
\end{equation}

\noindent
To measure the size of a query, we use the multiset containing the
level of each atom whose predicate is in $[p]_\approx$. The multiset is
formalised as a function $Size$, which takes as arguments a query and
a natural number:
\[
Size(Q)(n) = 
\# \{q(\vect{u},\vect{v}) \mid
q(\vect{u},\vect{v}) \in Q,\  
q \approx p \ \mbox{and} \
||q(\vect{u},\vect{v})|| = n \}.
\]
Note that 
if a query contains several identical atoms, each occurrence
must be counted.
We define $Size(Q) < Size(R)$ if and only if there is 
$l \in \nat$ such that
$Size(Q)(l) < Size(R)(l)$ and 
$Size(Q)(l') = Size(R)(l')$
for all $l' > l$.
Intuitively, there is a decrease when an atom
in a query is replaced with a finite number of smaller atoms.
All descending chains with respect
to $<$ are finite~\cite{D87}. 

Let $Q_0 = p(\vect{s},\vect{t})$ be a robustly typed query.  Then
$p(\vect{s},\vect{t}) \in \iatoms$ 
and thus $||Q_0||$ is defined.  Let 
$\xi = Q_0; Q_1;\dots$ be a delay-respecting derivation of 
$P \cup \{ Q_0 \}$.  

Since all predicates $q$ with $p \sqsupset q$ are robust, 
it follows by Lemma~\ref{robust=finite} that there cannot be an infinite suffix of
$\xi$ without any steps where an atom $q(\vect{u},\vect{v})$ such that
$q \approx p$ is resolved. We show that for all $i \geq 0$, 
if the selected atom in $Q_i \step Q_{i+1}$ is $q(\vect{u},\vect{v})$
and $q \approx p$, then $Size(Q_{i+1}) < Size(Q_i)$, and otherwise 
$Size(Q_{i+1}) \leq Size(Q_i)$.
This implies that $\xi$ is finite, and, as the choice of the initial query 
$Q_0 = p(\vect{s},\vect{t})$ was arbitrary, $p$ is robust.

By Lemma~\ref{robustly-typed-persistence}, each position in each atom in
$Q_{i+1}$ is filled with a type-consistent term.\hfill$(*)$

Consider $i \geq 0$ and let 
$C = q(\vect{v}_0,\vect{u}_{m+1}) \leftarrow
q_1(\vect{u}_1,\vect{v}_1),\dots,q_m(\vect{u}_m,\vect{v}_m)$
be the clause,
$q(\vect{u},\vect{v})$ the selected atom and 
$\theta$ the MGU used in $Q_i;Q_{i+1}$.

If $p \sqsupset q$, then 
$p \sqsupset q_j$ for all $j \in \onetom$, and hence by
(\ref{only-smaller}) and $(*)$ it follows that
$Size(Q_{i+1}) \leq Size(Q_i)$. Intuitively, the set of atoms that are
measured by $Size$ does not change in this step (although the level of
each atom might decrease).

Now consider $q \approx p$. Since $C$ is well-recurrent and because
of $(*)$, we have 
$|| q(\vect{v}_0,\vect{u}_{m+1}) \theta || >
 || q_j(\vect{u}_j,\vect{v}_j) \theta ||$ for all $j$ with 
$q_j \approx p$. This together with (\ref{only-smaller}) implies
$Size(Q_{i+1}) < Size(Q_i)$. Intuitively, one atom has been replaced
by smaller atoms in this step, but apart from that, the set of atoms 
that are measured by $Size$ does not change.
\end{proof} 

\noindent
{\bf Theorem~\ref{loop-relation-2}}
Let $P$ and $Q$ be a well fed program and query, 
and $P'$ and $Q'$ a robustly typed program and query
corresponding 
to $P$ and $Q$.
If every LD-derivation of $P' \cup \{ Q' \}$
is finite, then
every left-based derivation of
$P \cup \{ Q \}$ is finite.

\begin{proof}
Suppose there is an infinite left-based derivation $\xi$ of
$P \cup \{ Q \}$. Then letting $Q_0 = Q$, 
$\theta_0 = \emptyset$, we can write
\[
\xi = 
\langle Q_0, \theta_0 \rangle 
\steps 
\langle R_1, \sigma_1 \rangle 
\step
\langle Q_1, \theta_1 \rangle
\steps 
\langle R_2, \sigma_2 \rangle 
\step
\langle Q_2, \theta_2 \rangle 
\dots
\] 
where $R_1,R_2,\dots$ are the queries in $\xi$
where a non-robust atom is selected. By Lemma~\ref{robust=finite},
there are infinitely many such queries. We derive a
contradiction. 

By Prop.~\ref{well-fed-persistence},
the non-robust atoms in each query in $\xi$ have only ancestors in
safe positions.
Thus by Cor.~\ref{placed=instantiated}, for each $i > 1$, where 
$R_i$ is $\rho_i$-robustly typed, 
the ${{\rho_i}^{-1}(1)}$'th atom in $R_i$ is
selected in 
$\langle R_i, \sigma_i \rangle 
\step
\langle Q_i, \theta_i \rangle$.

Now consider an arbitrary query 
$\tilde{Q}$ 
in $\xi$ and assume it is $\tilde{\pi}$-robustly typed. 
By Cor.~\ref{mode-order} and the
previous paragraph it follows that there exists a query in $\xi$ that
contains no descendants of the 
${\tilde{\pi}^{-1}(1)}$'th atom $\tilde{Q}$. Intuitively, for each query
in $\xi$, the atom that is ``leftmost according to its permutation''
will eventually be resolved completely.

By repeatedly applying the Switching Lemma to
prefixes of $\xi$, we can construct a derivation $\zeta$ of
$P \cup \{ Q \}$ such that in each query 
$\tilde{Q}$ 
in $\zeta$ that is $\tilde{\pi}$-robustly typed, the 
$\tilde{\pi}^{-1}(1)$'th atom is selected using the same clause (copy) used in
$\xi$. Note that this construction is possible by the previous
paragraph. Also note that $\zeta$ is infinite.

Now consider the derivation $\zeta'$ obtained from $\zeta$ by replacing
each $\tilde{\pi}$-robustly typed query $\tilde{Q}$ with 
$\tilde{\pi}(\tilde{Q})$, i.e.~the robustly typed query corresponding
to $\tilde{Q}$. The derivation $\zeta'$ is an LD-derivation
of  $P' \cup \{ Q' \}$, and it is infinite. This is a contradiction.
 \end{proof}

\bibliography{thesis}

\end{document}